\documentclass[useAMS,usenatbib]{mn2e}
\usepackage[utf8x]{inputenc}
\usepackage{amssymb,amsmath,multirow}
\usepackage{comment,graphicx,overpic,natbib,apjfonts}
\usepackage{color,transparent}

\newcommand{\propsiminn}[2]{\mathrel{\vcenter{
  \offinterlineskip\halign{\hfil$##$\cr
    #1\propto\cr\noalign{\kern2pt}#1\sim\cr\noalign{\kern-2pt}}}}}
\newcommand{\propsim}{\mathpalette\propsiminn\relax}

\renewcommand{\vec}[1]{\boldsymbol{#1}}
\newcommand{\mr}[1]{\mathrm{#1}}

\newcommand{\eq}[1]{equation~\eqref{#1}}
\newcommand{\eqs}[2]{equations~\eqref{#1} \& \eqref{#2}}
\newcommand{\eqp}[1]{equation~\ref{#1}}
\newcommand{\eqsp}[2]{equations~\ref{#1} \& \ref{#2}}
\newcommand{\Eq}[1]{Equation~\eqref{#1}}

\newcommand{\fig}[1]{Figure~\ref{#1}}
\newcommand{\se}[1]{\S~\ref{#1}}
\newcommand{\app}[1]{Appendix~\ref{#1}}

\newcommand{\ad}{\mr{ad}}

\newcommand{\f}{\mr{f}}
\newcommand{\sign}{\mr{sign}}
\renewcommand{\Im}{\mr{Im}}
\renewcommand{\Re}{\mr{Re}}

\newcommand{\sigf}{\sigma_\f}
\newcommand{\Qf}{\Q_\f}
\newcommand{\ReQf}{\Re(\Qf)}
\newcommand{\ImQf}{\Im(\Qf)}
\newcommand{\domeg}{\delta\omega}
\newcommand{\domegf}{\domeg_\f}
\newcommand{\domegad}{\domeg_\ad}

\newcommand{\q}{q}

\newcommand{\G}{\Gamma}
\newcommand{\Gn}{\G_\tide}
\newcommand{\Gdr}{\G_\mr{dr}}

\newcommand{\Q}{Q}
\newcommand{\Qol}{\mathcal{I}}
\newcommand{\Qeq}{\mathcal{Q}_\mr{eq}}

\newcommand{\Es}{E_*}
\newcommand{\Is}{I_*}
\newcommand{\omegs}{\omega_*}
\newcommand{\betas}{\beta_*}

\newcommand{\orb}{\mr{orb}}
\newcommand{\spin}{\mr{spin}}
\newcommand{\mode}{\mr{mode}}
\newcommand{\tide}{\mr{tide}}

\newcommand{\atide}{\mathbf{a}_\tide}

\newcommand{\tgw}{t_\mr{gw}}

\renewcommand{\epsilon}{\varepsilon}

\newcommand\altaffilmark[1]{$^{#1}$}
\newcommand\altaffiltext[1]{$^{#1}$}

\title[Dynamical resonance locking]{Dynamical resonance locking in tidally interacting binary systems}

\author[Burkart et al.]{
\parbox[t]{\textwidth}{
Joshua Burkart,\altaffilmark{1}
Eliot Quataert,\altaffilmark{1,2}
Phil Arras\altaffilmark{3}}
\vspace*{6pt} \\
\altaffiltext{1}{Department of Physics, University of California,
Berkeley, CA}\\
\altaffiltext{2}{Department of Astronomy \& Theoretical Astrophysics
  Center, University of California, Berkeley, CA}\\
\altaffiltext{3}{Department of Astronomy, University of Virginia, Charlottesville, VA}
}

\begin{document}

\maketitle

\begin{abstract}
We examine the dynamics of resonance locking in detached, tidally interacting binary systems. In a resonance lock, a given stellar or planetary mode is trapped in a highly resonant state for an extended period of time, during which the spin and orbital frequencies vary in concert to maintain the resonance. This phenomenon is qualitatively similar to resonance capture in planetary dynamics. We show that resonance locks can accelerate the course of tidal evolution in eccentric systems and also efficiently couple spin and orbital evolution in circular binaries. Previous analyses of resonance locking have not treated the mode amplitude as a fully dynamical variable, but rather assumed the adiabatic (i.e.\ Lorentzian) approximation valid only in the limit of relatively strong mode damping. We relax this approximation, analytically derive conditions under which the fixed point associated with resonance locking is stable, and further check these analytic results using numerical integrations of the coupled mode, spin, and orbital evolution equations. These show that resonance locking can sometimes take the form of complex limit cycles or even chaotic trajectories. We provide simple analytic formulae that define the binary and mode parameter regimes in which resonance locks of some kind occur (stable, limit cycle, or chaotic). We briefly discuss the astrophysical implications of our results for white dwarf and neutron star binaries as well as eccentric stellar binaries.
\end{abstract}


\section{Introduction}\label{s:intro}
The most frequent treatment of tidal effects in detached binaries relies on the weak-friction theory \citep{murray}, which considers only the large-scale ``equilibrium tide'' (i.e., filling of the Roche potential) with dissipation parameterized by the tidal quality factor $\mathcal{Q}$ \citep{goldreich66}. Such a treatment fails to account for the resonant excitation of internal stellar waves with intrinsic frequencies comparable to the tidal forcing---the so-called ``dynamical tide'' \citep{zahn75}.

Many studies have considered the dynamical tide in different astrophysical contexts. There are two possible regimes: when the character of the excited modes is that of radially traveling waves, or when they represent standing waves. A traveling wave occurs when reflection is prohibited by a strong dissipative process at some radial location in the star or planet in question---e.g., rapid linear dissipation near the surface or nonlinear wave breaking.

The standing wave regime is the subject of this paper. This is applicable when a wave's amplitude can be built up by many reflections. Existing calculations in this regime have used at least one of the following two approximations: 1) tides do not backreact on the spin of the star or planet in question \citep{lai94,rathore05,fuller11}, or 2) the mode amplitude is not treated as a dynamical variable, and instead has its amplitude set by the adiabatic approximation discussed in \se{s:sec} \citep{witte99,fullerkoi,burkart13}.

In this paper we are interested in understanding the phenomenon of resonance locking, in which the orbital and spin frequencies vary in concert so as to hold the Doppler-shifted tidal forcing frequency $k\Omega_\orb-m\Omega_\spin$ constant \citep{witte99}. Resonance locking is analogous to the phenomenon of capture into resonance in planetary dynamics \citep{goldreich65,goldreich68}; we provide a comparison in \se{s:conc}. Resonance locks can accelerate the course of tidal evolution, as we will show in \se{s:evolve}. Moreover, recent studies \citep{fullerkoi,burkart12} have proposed that resonance locks may have been observed in the \emph{Kepler} system KOI-54 \citep{welsh11}, although this has been contested by \citet{oleary13}.

Since resonance locking involves a changing spin frequency, clearly it cannot occur under approximation (1) noted above. The domain of validity of approximation (2) is given in \se{s:sec} (see also \citealt{burkart13}). In this paper, we drop both of the above assumptions and examine resonance locks accounting for a dynamically evolving mode amplitude coupled to both the orbital and spin evolution. Our aim is to investigate the general dynamical properties of resonance locking, rather than to focus on a specific astrophysical application. Our key questions concern determining when resonance locks can occur and under what conditions they are dynamically stable.

This paper is structured as follows. We first describe the essential idea behind resonance locking in \se{s:basic}, and enumerate the approximations we make in order to limit the complexity of our analysis in \se{s:assump}. We then develop evolution equations for a single stellar or planetary eigenmode in \se{s:modeamp}, and determine the implied backreaction upon the binary orbit and stellar or planetary spin in \se{s:freq}. We establish the existence and assess the stability of fixed points in the evolution equations associated with resonance locking in \se{s:rcap}. We describe two analytic approximations that a mode's amplitude follows in certain limits in \se{s:analytic}, and then present example numerical integrations of resonance locks in \se{s:trajnum}. In \se{s:chaos} we discuss the possibility of chaos during resonance locking. We numerically and analytically determine the parameter regimes that lead to resonance locking in \se{s:achieve}, and show that resonance locks can accelerate tidal evolution in \se{s:evolve}. We apply our results to two example astrophysical systems---inspiraling compact object binaries and eccentric stellar binaries---in \se{s:astroph}. We then conclude in \se{s:conc}.

\section{Basic idea}\label{s:basic}
We first explain the essential mechanism behind resonance locking by considering the example situation of a circular white dwarf binary inspiraling due to the emission of gravitational waves \citep{burkart13}. Focusing on a particular white dwarf, and shifting to a frame of reference corotating with the white dwarf's spin, the tidal forcing frequency is $\sigma=m(\Omega_\orb - \Omega_\spin)$, where $m$ is the azimuthal spherical harmonic index and we temporarily assume $\Omega_\orb\gg\Omega_\spin$. Due to the influence of gravitational waves, $\Omega_\orb$ gradually increases, and thus so does $\sigma$. As such, $\sigma$ sweeps towards resonance with the nearest normal mode, and this mode gains energy as it becomes increasingly resonant \citep{rathore05}.

\begin{figure}\centering
  \begin{overpic}{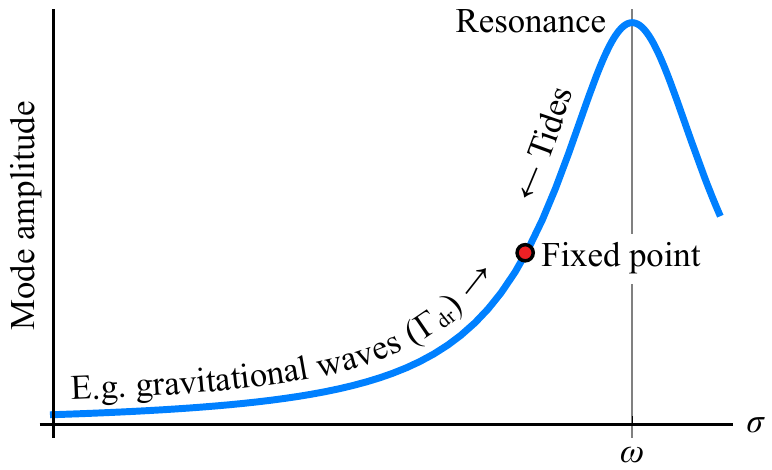}
  \end{overpic}
  \caption{Diagram illustrating the basic mechanism behind tidal resonance locking. Forcing frequency drift caused for example by gravitational waves causes the tidal forcing frequency $\sigma$ to advance towards the right. The influence of tidal synchronization, on the other hand, becomes stronger as resonance approaches, and tends to push the forcing frequency to the left. This creates a fixed point on the resonant mode's Lorentzian amplitude profile. (The mirror image of this situation is also possible.)}
  \label{f:basic}
\end{figure}

Along with energy, however, comes angular momentum (for $m\ne0$ modes). As the mode then damps, this angular momentum is transferred to the background rotation, increasing $\Omega_\spin$ and consequently decreasing $\sigma$. Thus if the mode is capable of achieving a sufficient amplitude, fixed points can exist where $\sigma$ (but not $\Omega_\orb$ or $\Omega_\spin$ individually) is held \emph{constant} by tidal synchronization balancing orbital decay by gravitational waves, as illustrated in \fig{f:basic}. This is the idea behind a tidal resonance lock.

The properties of such fixed points corresponding to resonance locking clearly depend on the rate of externally driven orbital evolution and the strength of tidal coupling to the mode in question. Furthermore, the mode's damping rate influences both its maximum achievable amplitude as well as the rate at which it dissipates angular momentum into the background rotation profile. Lastly, since resonance locking involves a balance between orbital and spin evolution, the ratio of the associated moments of inertia---roughly $MR^2 / \mu a^2$ where $\mu$ is the reduced mass and $a$ is the semi-major axis---also plays a key role. We will see in subsequent sections that dimensionless parameters corresponding to these four quantities will entirely determine the dynamics of resonance locking.

The ideas we have presented here in the context of inspiraling white dwarf binaries carry over to the more general scenario where some generic physical process causing the forcing frequency $\sigma$ or the eigenmode frequency $\omega$ to evolve in one direction---e.g., gravitational waves, magnetic braking, the equilibrium tide, stellar evolution, etc.---is balanced by the resonant tidal interaction with a planetary or stellar normal mode causing $\sigma$ to evolve in the opposite direction.

\section{Essential assumptions}\label{s:assump}
Throughout this work, we invoke the following principal assumptions.

\begin{enumerate}
\item As discussed in \se{s:intro}, we assume that the dynamical tide is composed of standing waves, and thus that linear dissipation (provided e.g.\ by radiative diffusion) is not strong enough to prohibit wave reflection.

\item As a mode's amplitude grows, nonlinear processes can become important. For example, gravity waves break and catastrophically dissipate when their vertical displacement becomes comparable to their vertical wavelength \citep{lindzen66}. Moreover, global parametric resonances can occur at smaller amplitudes, which transfer energy to a pair of daughter modes (e.g.\ \citealt{weinberg12}).

We completely neglect nonlinear hydrodynamical phenomena in this work. The applicability of this approximation depends on the particular star or planet's structure and the binary's parameters. For example, although \citet{witte02} considered resonance locking in solar-type binaries, \citet{goodman98} showed that the lack of a convective core in such stars allows geometric focusing to progressively amplify the local amplitude of an inward-propagating gravity wave to the point where wave breaking occurs over a wide range of binary parameters (see also \citealt{barker10}). This precludes the establishment of the standing waves required for a resonance lock.

\item We assume that the star or planet in question rotates as a rigid body, even though tidal torques are typically applied to the background rotation profile in a thin spherical shell where mode damping is strongest \citep{goldreich89}. We thus rely upon the action of an efficient angular momentum transport process. Whether such a process is available strongly depends on the application in question; see e.g.\ \citet{burkart13} and accompanying references. Without solid-body rotation, critical layers can develop where angular momentum is deposited, which would provide rapid, local dissipation and hence violate assumption (i).

\item We take the overall strength of a mode's tidal forcing to be constant, and account for orbital evolution only insofar as it affects the forcing \emph{frequency}. We thus ignore, for example, that the magnitude of the tidal force depends on the binary's semi-major axis, which necessarily must evolve if $\Omega_\orb$ is to change. We determine when this assumption is valid in \se{s:exist}; it would never be appropriate for long-timescale simulations that follow the evolution of many resonance locks (which are not performed in this paper).

\item Lastly, to simplify our analysis, we only consider binary systems where the spin and orbital angular momenta are aligned. Generalization to spin-orbit-misaligned systems is straightforward \citep{lai12}, and ultimately only introduces a Wigner $\mathcal{D}$-matrix element into our definition of the tidal coupling coefficient $U$ in \se{s:modeamp}.
\end{enumerate}

\section{Formalism}\label{s:form}
\subsection{Mode amplitude evolution}\label{s:modeamp}
We work in a frame of reference corotating with the spin of the body in question, where the rotation axis lies along the $\hat{z}$ direction. Following \citet{schenk02}, we invoke a phase space eigenmode decomposition of the tidal response \citep{dyson79}. The standard stellar mode inner product for arbitary vector fields $\vec{\eta}$ and $\vec{\eta}'$ is
\begin{equation}
 \langle\vec{\eta},\vec{\eta}'\rangle = \int \vec{\eta}^*\cdot\vec{\eta}' \rho\, dV,
\end{equation}
and the anti-Hermitian Coriolis force operator is defined by $B\,\vec{\eta} = 2\Omega_\spin\hat{z}\times\vec{\eta}$.

We consider the resonant interaction between a time-varying tidal potential evaluated on a Keplerian orbit and a single linear mode of our expansion. The mode has spherical harmonic indices $l$ and $m$, and has complex amplitude $q$. The differential equation describing the mode's linear evolution is \citep{schenk02}
\begin{equation}\label{e:modeamp}
   \dot{q} + (i\omega+\gamma) q = \frac{i\omega}{\epsilon}\left\langle\vec{\xi}, \atide\right\rangle,
\end{equation}
where $\omega$ is the mode's eigenfrequency, $\gamma$ is its damping rate, $\vec{\xi}$ is its Lagrangian displacement vector,
\begin{equation}
 \epsilon = 2\omega^2\langle\vec{\xi},\vec{\xi}\rangle + \omega\langle\vec{\xi},iB\vec{\xi}\rangle
\end{equation}
is a normalization factor (equal to the mode energy at unit amplitude), and $\atide$ is the time-dependent tidal acceleration vector. We assume $\gamma\ll\omega$ and, without loss of generality, we take $\omega>0$.

Allowing for an arbitrary eccentricity, the projection of the tidal acceleration onto the mode $\left\langle\vec{\xi}, \atide\right\rangle$ is proportional to
\begin{displaymath}
\left(\frac{a}{D}\right)^{l+1}e^{-im\left(f-\psi_\spin\right)},
\end{displaymath}
where $D$ is the binary separation, $a$ is the semi-major axis, $f$ is the true anomaly, and
\begin{equation}\label{e:phases}
 \begin{pmatrix}\psi_\orb\\\psi_\spin\end{pmatrix} = \int \begin{pmatrix}\Omega_\orb\\\Omega_\spin\end{pmatrix} dt.
\end{equation}
Assuming that changes in the orbital frequency and eccentricity occur on timescales much longer than an orbital period, we can expand the dependence on $D$ and $f$ in a Fourier series:
\begin{equation}
 \left(\frac{a}{D}\right)^{l+1}e^{-im\left(f-\psi_\spin\right)} \approx \sum_k X_{lm}^k e^{-i\left(k\psi_\orb-m\psi_\spin\right)},
\end{equation}
where $X_{lm}^k(e)$ is a Hansen coefficient (\app{a:hansen}), with $X_{lm}^k(e=0)=\delta^k_{m}$ for circular orbits. Since we are concerned with resonant mode-tide interaction, we henceforth consider only a single harmonic component of this expansion. The phase associated with this component is $\psi = k \psi_\orb-m\psi_\spin$.

Our mode amplitude equation from \eqref{e:modeamp} then becomes
\begin{equation}\label{e:modeamp2}
 \dot{q} + (i\omega+\gamma)q = i\omega U e^{-i\psi}.
\end{equation}
The dimensionless mode-tide coupling strength associated with our harmonic component is
\begin{equation}\label{e:U}
U = \left( \frac{M'}{M} \right)\left( \frac{R}{a} \right)^{l+1}\left( \frac{\Es}{\epsilon} \right)\Qol_{lm}\, X_{lm}^k\, W_{lm},
\end{equation}
where $M'$ is the companion mass,
\begin{equation}
 \Qol_{lm} = \frac{1}{MR^l}\left\langle\vec{\xi},\vec{\nabla}\left(r^l Y_{lm}\right)\right\rangle
\end{equation}
is the mode's linear tidal overlap integral \citep{press77},
\begin{equation}
 W_{lm} = \frac{4\pi}{2l+1}Y_{lm}^*\left( \frac{\pi}{2},\,0 \right)
\end{equation}
is an order-unity angular coupling coefficient, and $\Es=GM^2/R$ is the gravitational energy scale. We also define the tidal forcing frequency to be $\sigma = \dot{\psi} = k\Omega_\orb - m\Omega_\spin$, and take $\sigma>0$ without loss of generality. We are free to choose the sign of $\sigma$ since we are considering a complex conjugate mode pair---one member of the pair is resonant with $\sigma$ while the other is resonant with $-\sigma$. Because $m$ can possess either sign, both prograde and retrograde modes are allowed for $\sigma>0$. Since we are considering resonant interaction, our assumption is $\sigma\approx\omega$.

\subsection{Forcing frequency evolution}\label{s:freq}
We will first obtain the time derivative of the stellar or planetary spin frequency implied by \eq{e:modeamp2}. The canonical angular momentum associated with the mode together with its complex conjugate is given by (\app{a:J})
\begin{equation}\label{e:Jmode}
 J_\mode = \frac{m\epsilon}{\omega}|q|^2.
\end{equation}
Differentiating with respect to time and substituting \eq{e:modeamp2}, we find
\begin{equation}\label{e:Jdotmode}
   \dot{J}_\mode = -2\gamma J_\mode + 2m\epsilon U \Im\left[ \q e^{i\psi} \right].
\end{equation}

The first term in \eq{e:Jdotmode} results from damping, and is imparted to the stellar or planetary spin \citep{goldreich89b}.\footnote{It can be shown that the second term in \eq{e:Jdotmode} is exactly the angular momentum transfer rate from the orbit; see e.g.\ \citet{weinberg12}. This justifies attributing the first term to changes in the spin.} We thus set
\begin{equation}
   \dot{J}_\spin = 2\gamma J_\mode,
\end{equation}
meaning that
\begin{equation}\label{e:Omegdotspin}
   \dot{\Omega}_\spin = \frac{2\gamma J_\mode}{\Is} + \alpha_\spin,
\end{equation}
where $\Is$ is the planet or star's moment of inertia, and we have incorporated an additional term $\alpha_\spin$ to account for processes that can change $\Omega_\spin$ other than interaction with the mode in question---e.g., the equilibrium tide, magnetic braking, etc. We take $\alpha_\spin$ to be constant.

The rate at which the orbital energy changes is given by \citep{weinberg12}
\begin{equation}\label{e:Edotorb}
   \dot{E}_\orb = -2k\Omega_\orb\epsilon U \Im\left[ \q e^{i\psi} \right].
\end{equation}
We can convert \eq{e:Edotorb} into an expression for $\dot{\Omega}_\orb$ using
\begin{equation}\label{e:EOmeg}
   \frac{\dot{\Omega}_\orb}{\Omega_\orb} = -3\,\frac{\dot{E}_\orb}{\mu a^2 \Omega_\orb^2 },
\end{equation}
which yields
\begin{equation}\label{e:Omegdotorb}
   \dot{\Omega}_\orb = \left(\frac{3k}{\mu a^2}\right)
   2\epsilon U\Im\left[ \q e^{i\psi} \right] + \alpha_\orb,
\end{equation}
where we have again included an extra term $\alpha_\orb$ to account for e.g.\ orbital decay by gravitational waves.

It is useful to combine \eqs{e:Omegdotspin}{e:Omegdotorb} to determine the time derivative of $\domeg=\omega-\sigma$, which can be expressed as
\begin{equation}\label{e:eomsig}
   \frac{\dot{\domeg}}{\omega} = -\Gdr + \Gn \left( \frac{\gamma|\q|^2}{\omega U^2} -r\, \Im\left[\frac{\q e^{i\psi}}{U}\right] \right).
\end{equation}
Here we have combined both $\alpha$ parameters from earlier into the ``drift'' rate
\begin{equation}\label{e:Gdr}
   \Gdr = \frac{k\alpha_\orb-m(1-C)\alpha_\spin}{\omega} - \frac{\partial \ln\omega}{\partial t},
\end{equation}
where
\begin{equation}
 C = -\frac{1}{m}\frac{\partial\omega}{\partial\Omega_\spin}
\end{equation}
allows for a rotationally dependent corotating-frame eigenmode frequency and $\partial\omega/\partial t$ accounts for changes in the eigenmode frequency due to progressive modifications of the background hydrostatic profile from e.g.\ stellar evolution.\footnote{Tidal heating could contribute to the $|q|^2$ term in \eq{e:eomsig}, due to heat deposited by the mode in question affecting the background star or planet; we neglect this for simplicity.} The tidal backreaction rate is
\begin{equation}\label{e:Gn}
   \Gn = \frac{2m^2 U^2 (1-C)\epsilon}{\Is \omega},
\end{equation}
and parameterizes the strength of tidal coupling to the mode in question; it is related to the rate at which the mode can synchronize the star or planet at nonresonant amplitudes ($|q|\sim |U|$). Lastly, the moment of inertia ratio is
\begin{equation}\label{e:r}
   r = \frac{k^2}{m^2}\frac{3\Is}{(1-C)\mu a^2}.
\end{equation}
We assume that both $\Gn$ and $r$ are positive throughout this work, but allow $\Gdr$ to possess either sign.

\section{Resonance lock fixed points}\label{s:rcap}
\subsection{Existence of fixed points}\label{s:exist}
We first remove the overall oscillatory time dependence of $q$ by changing variables to $\Q=\q e^{i\psi}/U$, so that \eq{e:modeamp2} becomes
\begin{equation}\label{e:eom1}
 \dot{\Q} + \left( i\domeg+\gamma +\frac{\dot{U}}{U}\right)\Q=i\omega,
\end{equation}
where again $\domeg=\omega-\sigma$. For simplicity, we now assume that $\gamma$ is much larger than terms contributing to $\dot{U}/U$, such as $\dot{a}/a$ and $\dot{e}/e$, so that \eq{e:eom1} becomes
\begin{equation}\label{e:eom2}
 \dot{\Q} + \left( i\domeg+\gamma\right)\Q=i\omega;
\end{equation}
as we will see in \se{s:evolve}, this essentially amounts to assuming $\gamma\gg|\Gdr|$. \Eq{e:eomsig} becomes
\begin{equation}\label{e:eomsig2}
 \frac{\dot{\domeg}}{\omega} = -\Gdr + \Gn\left( \frac{\gamma}{\omega}|\Q|^2 - r\,\Im[\Q] \right).
\end{equation}
Having thus eliminated direct dependence on the phase $\psi$, our two dynamical variables are now $\Q$ and $\domeg$. Since $Q$ is complex, we have a third-order differential system.

Resonance locking corresponds to a fixed point in the evolution equations. We thus set time derivatives to zero in \eq{e:eom2} to derive
\begin{equation} \label{e:Qf}
 \Qf = \frac{\omega}{\domegf - i\gamma},
\end{equation}
and hence
\begin{equation}\label{e:sol1}
 \begin{bmatrix}
  \ReQf\\\ImQf
 \end{bmatrix} = \frac{\omega}{\domegf^2 + \gamma^2}\times
 \begin{bmatrix}
  \domegf\\\gamma
 \end{bmatrix}.
\end{equation}
Similarly, setting $\dot{\domeg}=0$ in \eq{e:eomsig2} and using the fact that $(\gamma/\omega)|\Qf|^2=\ImQf$, we have
\begin{equation}\label{e:sol2}
 \Gdr = (1-r)\,\Gn\,\ImQf.
\end{equation}
We can use \eqs{e:sol1}{e:sol2} to derive
\begin{equation}\label{e:domeg}
 \domegf^2 = \frac{\omega\gamma\Gn(1-r)}{\Gdr}-\gamma^2.
\end{equation}
So far we have not determined the sign of $\domegf$, and indeed there is one fixed point for each sign; however, we will show in \se{s:eig} that one is always unstable.

These fixed points exists if \eqs{e:sol1}{e:sol2} can be solved for $\Qf$ and $\sigf$, which is possible if (assuming $\Gn>0$)
\begin{equation}\label{e:exist1}
 \frac{\gamma}{\omega\Gn} < \frac{1-r}{\Gdr};
\end{equation}
this in particular requires
\begin{equation}\label{e:exist2}
(1-r)\Gdr>0.
\end{equation}
If $\Gdr>0$, then \eq{e:exist2} reduces to
\begin{equation}\label{e:critharm}
   |k|<|m|\left(\frac{(1-C)\mu a^2}{3\Is}\right)^{1/2}.
\end{equation}
As a result, \citet{fullerkoi} referred to the quantity on the right-hand side of \eq{e:critharm} as the ``critical'' harmonic.

\label{s:weaktide} The existence of the fixed points also relies on the ``weak-tide'' limit, which we define to be $|\domegf|<\Delta\omega/2$, where $\Delta\omega$ is the eigenmode frequency spacing near the mode in question. This means that the fixed point must lie within the eigenmode's domain of influence, so that the contribution of other eigenmode resonances can be legitimately neglected. If this is not the case, then resonance locking is not possible.

For example, if $\Gdr$ is very small, \eq{e:sol2} requires a commensurately small value of $Q$. However, this could be impossible to achieve in practice, since it would require a very large value of the detuning $\domeg$, allowing the possibility for a neighboring mode to come into resonance. The actual outcome in such a situation is that the tidal interaction would dominate the dynamics, and the drift processes contributing to $\Gdr$ would be irrelevant---i.e., the ``strong-tide'' limit.

A necessary condition for the weak-tide limit, and thus for a resonance lock to be able to occur, is $|\domegf|\ll\omega$, which evaluates to
\begin{equation}\label{e:weaktide}
   \frac{|\Gdr|}{\gamma} \gg \frac{\Gn|1-r|}{\omega}.
\end{equation}
We will use this as a convenient, although very liberal, proxy for the real requirement of $|\domegf|<\Delta\omega/2$, so as to avoid handling the additional parameter $\Delta\omega$. Calculations of dynamical tidal evolution that use the adiabatic approximation (\se{s:sec}) with many modes---e.g.\ \citet{witte99,fullerkoi}---already naturally account for the true requirement.

\subsection{Fixed point stability}\label{s:stab}
\subsubsection{Linearization}
We will now perform a linear stability analysis about each fixed point. First, note that the presence of the nonanalytic functions $|\cdot|$ and $\Im(\cdot)$ in \eq{e:eomsig2} necessitates treating the real and imaginary parts of $\Q$ separately. Thus let
\begin{displaymath}
\vec{\zeta} = \begin{bmatrix}\Re(\Q)&  \Im(\Q)& (\sigma-\omega)/\omega\end{bmatrix}^T,
\end{displaymath}
and
\begin{displaymath}
 \frac{d\vec{\zeta}}{dt} = \mathbf{f}(\vec{\zeta}),
\end{displaymath}
where $\mathbf{f}$ represents \eqs{e:eom2}{e:eomsig2} and satisfies $\vec{\nabla} \cdot \mathbf{f} = -2\gamma$. We can then derive a time-evolution equation for $\vec{\delta\zeta} = \vec{\zeta} - \vec{\zeta}_\f$ by invoking results from \se{s:exist}.

Proceeding, to linear order we have
\begin{equation}
 \frac{d}{dt}\vec{\delta\zeta} = A \vec{\delta\zeta} + o(\vec{\delta\zeta}),
\end{equation}
where
\begin{equation}\label{e:A}
 A=\mathcal{D}\mathbf{f}(\vec{\zeta}_\f)=\begin{bmatrix}
    -\gamma & \domegf & A_{1,3} \\
    -\domegf & -\gamma & A_{2,3} \\
    A_{3,1} & A_{3,2} & 0
   \end{bmatrix},
\end{equation}
\begin{align}
 &&A_{1,3} &= -\frac{\omega\beta}{\Gn} &
 A_{2,3} &= \frac{\omega\beta\domegf}{\gamma\Gn}&&\\
 &&A_{3,1} &= -\frac{2\beta\domegf}{\omega} &
 A_{3,2} &= r\Gn - \frac{2 \beta \gamma}{\omega},&&
\end{align}
and $\beta = \Gdr/(1-r)$.

\subsubsection{Eigenvalues}\label{s:eig}
We assume in this section that \eq{e:exist1} is satisfied, so that the resonance locking fixed point formally exists, and that \eq{e:weaktide} is also satisfied, so that we are in the weak-tide limit and the fixed point physically exists. The characteristic polynomial $P$ for the eigenvalues $\lambda$ of $A$ is
\begin{equation}\label{e:char}
   P(\lambda)=\lambda^3 + 2\gamma\lambda^2+P_1\lambda+P_0,
\end{equation}
where
\begin{equation}
   P_1 =  \frac{\omega\gamma\Gn}{\beta} - \frac{r\omega\beta\domegf}{\gamma} \qquad P_0 = 2\omega\Gdr\domegf.
\end{equation}
A fixed point is asymptotically stable if all eigenvalues satisfy $\Re(\lambda)<0$. A standard theorem then states that, for this to occur, it is necessary (but not sufficient) that all of the coefficients of $\lambda^i$ ($i\ge0$) in \eq{e:char} possess the same sign. We thus must have that $P_{1,0}>0$.

First, we see that
\begin{equation}\label{e:rightfp}
+1=\sign(P_0)=\sign(\Gdr)\sign(\domegf).
\end{equation}
Since equations \eqref{e:sol1} \& \eqref{e:sol2} admit two solutions, corresponding to $\domegf=\pm|\domegf|$, the criterion that $P_0>0$ simply allows us to select the solution that could potentially be stable. Henceforth we will focus on the solution that satisfies \eq{e:rightfp}, which we refer to as the ``lagging'' fixed point; similarly, the ``leading'' fixed point is the one that fails to satisfy \eq{e:rightfp}.

\Eq{e:rightfp} permits the following intuitive interpretation: the forcing frequency $\sigma$ is ``pushed'' towards an eigenfrequency in the direction specified by the sign of $\Gdr$, but the approaching eigenmode ``pushes'' $\sigma$ in the opposite direction, and resonance locking occurs when these ``forces'' cancel out (\se{s:basic}). In particular, if $\Gdr>0$, $\sigf$ should be smaller than $\omega$, meaning $\domegf=\omega-\sigf>0$, consistent with \eq{e:rightfp}.

It remains to analyze $P_1$ and to ascertain when it is also positive. Moreover, since positivity of the characteristic polynomial's coefficients is only a necessary condition for stability, we must then further examine the Hurwitz matrix associated with $P$ to establish when its leading principal minors are also positive (e.g.\ \citealt{gradshteyn}). We perform this analysis in \app{a:fixstab} in the limit of $\gamma\ll|\domegf|$; the result is that the lagging fixed point is stable if and only if
\begin{equation}\label{e:estabul}
   \Gdr<0 \quad \text{or} \quad 0<\frac{\Gdr}{\gamma}<(1-r)\left(\frac{\Gn}{\omega}\right)^{1/3}
\end{equation}
(subject to the assumptions we have made thus far). In particular, the lagging fixed point is thus always stable for $r>1$ per \eq{e:exist2}.

\begin{figure}
  \begin{overpic}{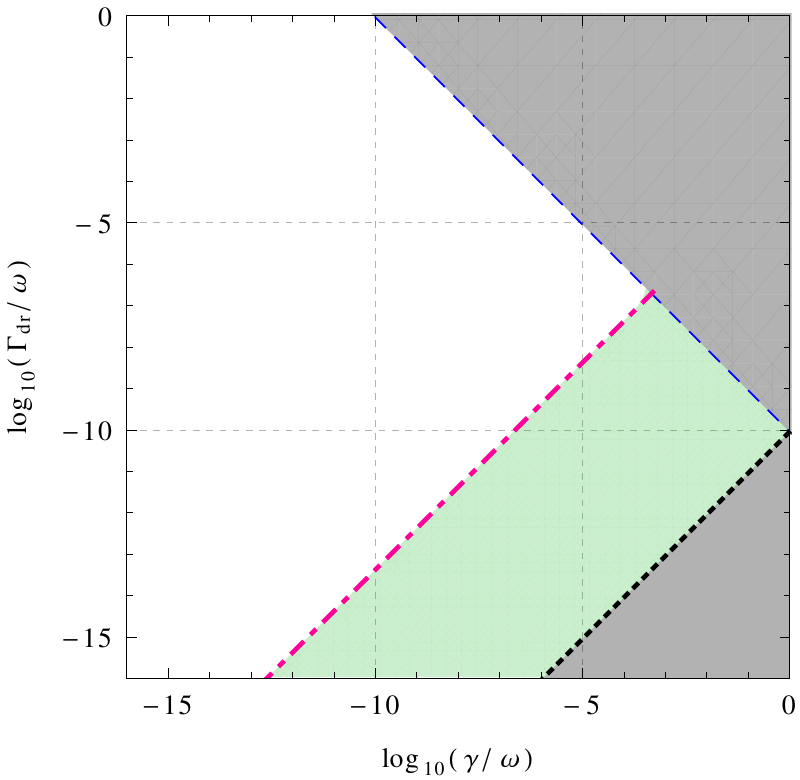}
  \put(33,63){Unstable}
  \put(72,80){Does not exist}
  \put(65,30){Stable}
  \put(81,20){Strong-tide}
  \put(81,16){regime}
  \put(17,90){\transparent{0}\colorbox{white}{\transparent{1}$\log_{10}(\Gn/\omega)=-10$}}
  \put(17,85){\transparent{0}\colorbox{white}{\transparent{1}$r=0.1$}}
  \end{overpic}\\[.2cm]
  \begin{overpic}{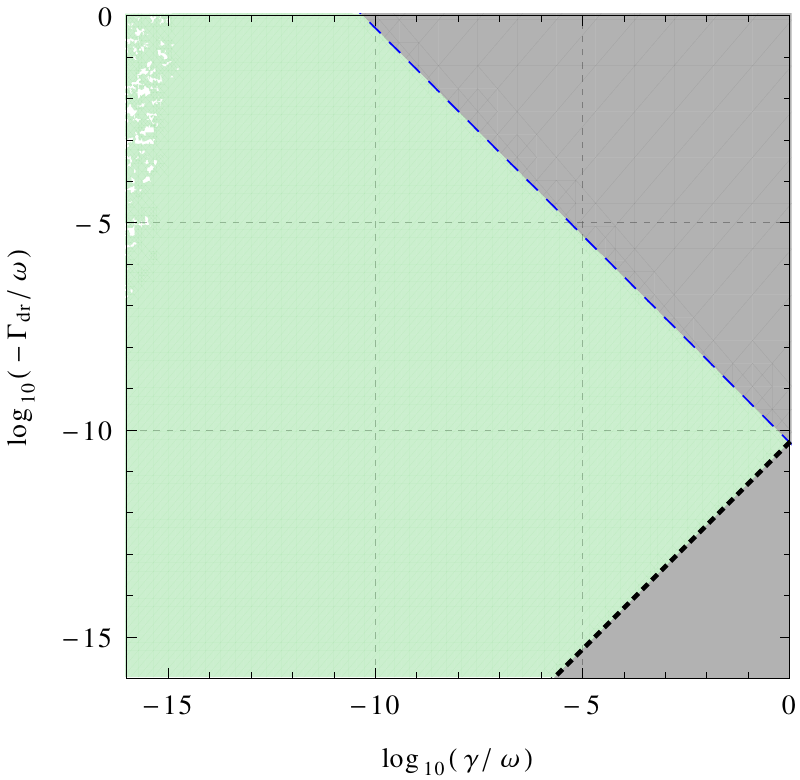}
  \put(48,45){Stable}
  \put(72,80){Does not exist}
  \put(81,20){Strong-tide}
  \put(81,16){regime}
  \put(17,90){\transparent{0}\colorbox{white}{\transparent{1}$\log_{10}(\Gn/\omega)=-10$}}
  \put(17,85){\transparent{0}\colorbox{white}{\transparent{1}$r=1.5$}}
  \end{overpic}
  \caption{Resonance locking fixed point stability analysis. Regions where the fixed point is asymptotically stable are shaded green, while unstable regions are white, and regions where the fixed point does not exist are dark gray. We determined the stability regions by numerically evaluating the eigenvalues $\lambda_i$ of the matrix $A$ determined by linearizing the equations of motion about the fixed point (\eqp{e:A}) and enforcing $\Re(\lambda_i)<0$ together with \eq{e:weaktide}. The analytic results in equations \eqref{e:exist1}, \eqref{e:weaktide}, \& \eqref{e:estabul} are displayed as blue dashed, black dotted, and magenta dot-dashed lines. The lower-right gray triangle in both panels corresponds to where the weak-tide limit is certainly violated, and thus the fixed point does not physically exist; see \se{s:exist}. In the bottom panel we have $r>1$, which by \eq{e:exist2} implies $\Gdr<0$.}
  \label{f:eigstab}
\end{figure}
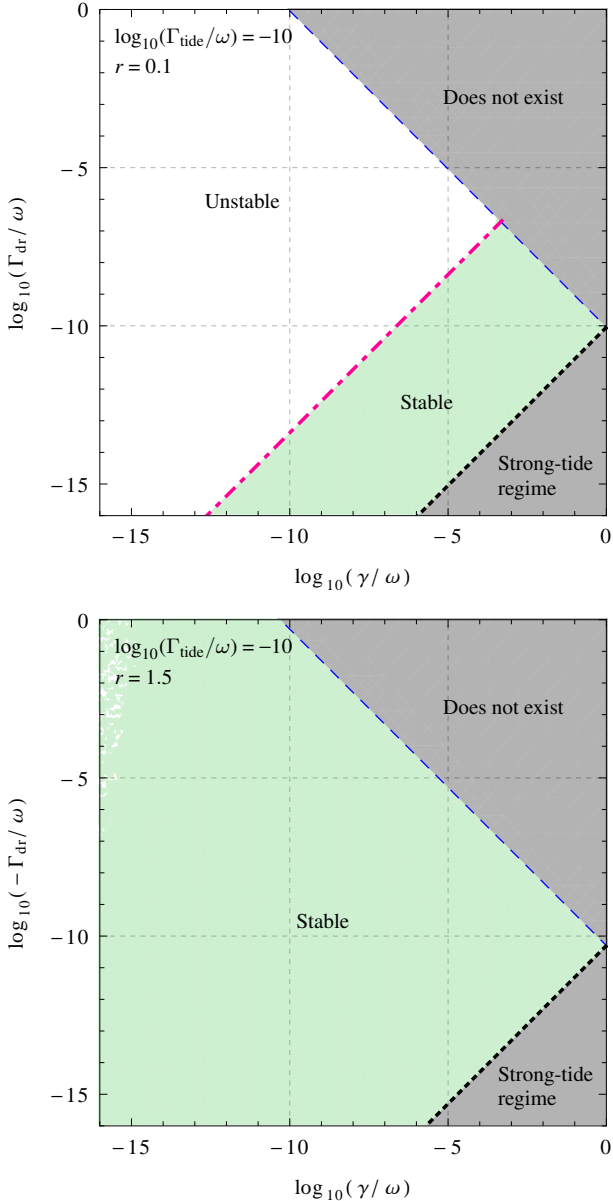

\fig{f:eigstab} shows the stability region of the lagging fixed point as a function of the damping rate $\gamma$ and the frequency drift rate $\Gdr$ for two example values of the backreaction rate $\Gn$ and the moment of inertia ratio $r$. Stability was determined by numerically solving for the eigenvalues $\lambda$ using \eq{e:char}, and is indicated by green shading, while instability is white; regions where the fixed point does not exist are shaded dark gray. Equations \eqref{e:exist1}, \eqref{e:weaktide}, \& \eqref{e:estabul}, which are displayed as blue dashed, black dotted, and magenta dot-dashed lines, closely correspond to the green region's boundaries. The values used in the top panel were chosen based on our white dwarf binary application in \se{s:wd}. In the bottom panel we have $r>1$, which by \eq{e:exist2} implies $\Gdr<0$.

\begin{figure}
  \begin{overpic}{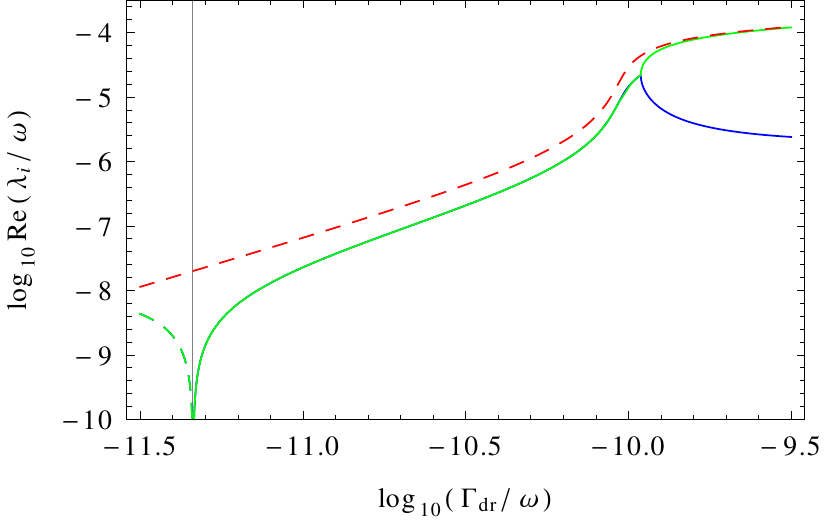}
  \put(64,17){Dashed indicates $<0$}
  \put(24,36){\rotatebox{90}{Hopf bifurcation}}
  \put(35,27){\rotatebox{20}{Complex conjugate pair}}
  \end{overpic}
  \caption{Real parts of the lagging fixed point's three eigenvalues as functions of $\Gdr$ for $\gamma/\omega=10^{-8}$, $\Gn/\omega=10^{-10}$, and $r=0.01$ (see the left middle panel of \fig{f:capregime}). A complex conjugate pair exists for $\Gdr/\omega\lesssim10^{-10}$ (\eqp{e:ccsplit}), which loses stability at $\Gdr/\omega\approx10^{-11.3}$ in a Hopf bifurcation (\eqp{e:estabul}).}
  \label{f:evals}
\end{figure}

The instability boundary defined by \eq{e:estabul} is in fact a supercritical Hopf bifurcation \citep{wiggins}, corresponding to the loss of stability of a complex conjugate eigenvalue pair. Past the bifurcation, inside the unstable region of parameter space, this unstable complex conjugate pair splits into two unstable real eigenvalues, as shown in \fig{f:evals}. In other words, the fixed point switches from being an unstable spiral to an unstable node. Determining the parameter space manifold on which this splitting occurs will be useful in \se{s:achieveanalyt}; we can accomplish this by setting the discriminant of \eq{e:char} to zero (thus requiring a repeated root), yielding
\begin{equation}\label{e:discriminant}
 0=36\gamma P_1 P_0 - 32 \gamma^3 P_0+4\gamma^2 P_1^2 - 4 P_1^3-27P_0^2.
\end{equation}
Although this equation cannot be solved analytically, we can nonetheless determine the asymptotic dependence of $\Gdr$ on $\gamma$ in the limit $\gamma\rightarrow0$. Newton's polygon for \eq{e:discriminant} shows that this asymptotic dependence is linear: $\Gdr = A \gamma$. Substituting this into \eq{e:discriminant}, setting $\gamma=0$, and solving for $A$, we have that the lagging fixed point is a spiral if
\begin{equation}\label{e:ccsplit}
 \Gdr < \gamma\left(\frac{1-r}{r^{2/3}}\right)\left( \frac{\Gn}{\omega} \right)^{1/3}.
\end{equation}

\section{Trajectories}\label{s:traj}
Here we will show several examples of trajectories that can be produced by our dynamical equations from \se{s:exist}. First, however, in \se{s:analytic} we will discuss two different analytic approximations that the trajectories follow in certain limits.

\subsection{Analytic approximations}\label{s:analytic}
\subsubsection{Adiabatic approximation}\label{s:sec}
In this paper, we define the adiabatic approximation to be the situation where the mode amplitude can instantaneously adjust to a changing forcing frequency, and thus we can set the $\dot{Q}$ term in \eq{e:eom2} to zero and assume $\sigma$ is constant. \Eq{e:eom2} can then be solved exactly:
\begin{equation}\label{e:sec}
 \Q_\ad = \frac{\omega}{\domeg-i\gamma}.
\end{equation}
This approximation, also referred to as the Lorentzian approximation due to the form of \eq{e:sec}, is frequently employed in the literature.

The domain of validity of the adiabatic approximation can be determined by comparing the maximum possible mode growth rate to the growth rate implied by \eq{e:sec}; when the latter exceeds the former, the adiabatic approximation is no longer valid. We can determine the maximum rate at which a mode amplitude $\Q$ can grow by providing a perfect resonance to \eq{e:eom2}, i.e., by setting $\sigma=\omega$. Dropping the damping term and setting $\dot{\sigma}=0$, the particular solution is
\begin{equation}
 \Q(t) =i \omega t,
\end{equation}
which implies that $|\dot{\Q}|_\mr{max}=\omega$. Next, in order to estimate the time derivative of $\Q_\ad$, we take $\dot{\sigma} \approx \Gdr\omega$, which gives us
\begin{equation}
 |\dot{\Q}_\ad| \approx |\Gdr|\left( \frac{\omega^2}{\domeg^2+\gamma^2}\right).
\end{equation}
We then equate $|\dot{\Q}_\ad|$ and $|\dot{\Q}|_\mr{max}$ and solve for $\domeg$, finding
\begin{equation}\label{e:secinv}
 \domeg_\ad^2 \approx |\Gdr|\omega-\gamma^2;
\end{equation}
if $|\Gdr|\omega<\gamma^2$, then the adiabatic approximation is always valid.

\subsubsection{No-backreaction approximation}\label{s:nbr}
Although directly solving \eqs{e:eom2}{e:eomsig2} outside the adiabatic limit requires numerical integration, we can nonetheless produce an approximate analytic expression for $\Q(t)$ in the limit that backreaction of the mode on the tidal forcing frequency $\sigma$ is unimportant \citep{reisenegger94,rathore05}. This approximation subsumes the adiabatic approximation, but is also more complicated.

Since we are already assuming that the mode damping rate is weak enough for the resonance locking fixed point to exist (\eqp{e:exist1}), we can simply take $\gamma\rightarrow0$. Subject to this simplification, we can solve \eq{e:modeamp2}, yielding
\begin{equation}\label{e:vop}
 \q(t) \approx i\omega Ue^{-i\omega t}\int_{t_0}^t e^{i\omega t-i\psi}dt,
\end{equation}
with $t_0\ll-(\omega\Gdr)^{-1/2}$. If we then approximate $\psi$ as
\begin{displaymath}
\psi\approx\psi_0+\omega t + \omega\Gdr t^2/2,
\end{displaymath}
where we have set resonance to occur (i.e.\ $\dot{\psi}=\omega$) at $t=0$, then \eq{e:vop} becomes a closed-form solution to the equations of motion.

Since the integral in \eq{e:vop} approaches a constant for $t\gg(\omega\Gdr)^{-1/2}$, and since we wish to estimate the maximum value of $|\Q|$, we can simply extend the domain of integration to $(-\infty,+\infty)$, yielding (using e.g.\ the method of stationary phase)
\begin{equation}
 \q(t) \approx -(1-i)U\sqrt{\frac{\pi\omega}{\Gdr}}e^{-i\omega t-i\psi_0}
\end{equation}
for $t\gg(\omega\Gdr)^{-1/2}$. We thus find that
\begin{equation}\label{e:Qmax}
 |\Q|_\mr{max}\approx \sqrt{\frac{2\pi\omega}{|\Gdr|}}.
\end{equation}
We plot this maximal value of $|\Q|$ in the middle panels of Figures \ref{f:trajplot1} -- \ref{f:trajplot4} as a dash-dotted black line.

\subsection{Numerical results}\label{s:trajnum}

\begin{figure}
  \includegraphics{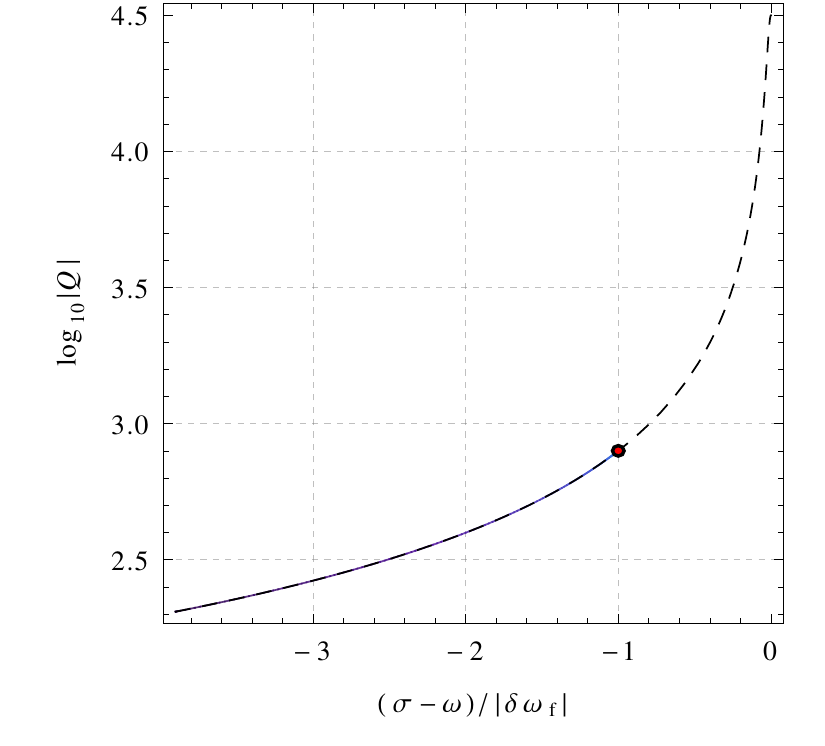}\\[.1cm]
  \includegraphics{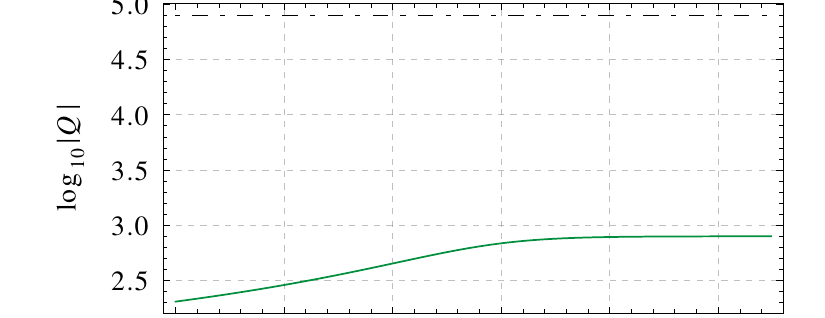}
  \includegraphics{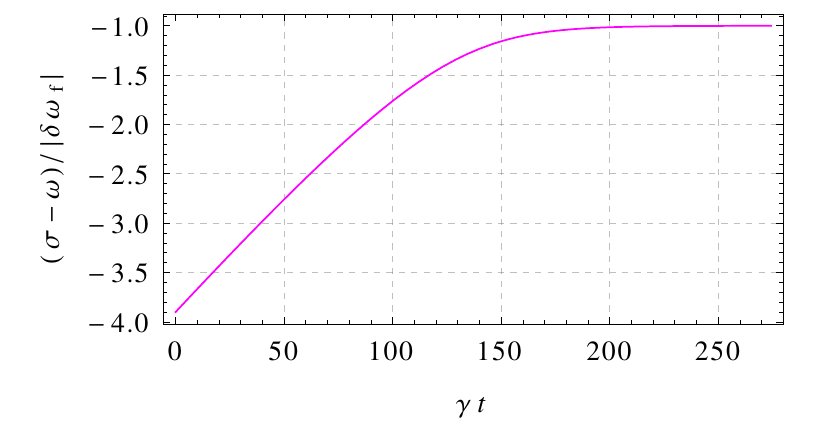}
  \caption{Numerical integration of mode and orbital evolution equations for the case of resonance locking into a stable fixed point. Parameters used were $\Gdr/\omega=10^{-9}$, $\gamma/\omega=10^{-4.5}$, $\Gn/\omega=10^{-10}$, and $r=0.5$. The adiabatic approximation is shown in the top panel as a dashed black line (\se{s:sec}), while the actual system trajectory is purple. The red circle shows the lagging  fixed point (\se{s:rcap}). Individual timeseries for the mode amplitude $\Q$ and the forcing frequency $\sigma$ are shown in the bottom two panels. The dash-dotted black line in the mode amplitude panel shows \eq{e:Qmax}, which gives the maximum amplitude attainable under the no-backreaction approximation (\se{s:nbr}). The lagging fixed point's eigenvalues are $\lambda/\omega\in\{(-0.031 \pm 1.2\,i)\times10^{-3},\ -1.6\times10^{-6}\}$.}
  \label{f:trajplot1}
\end{figure}

\begin{figure}
  \includegraphics{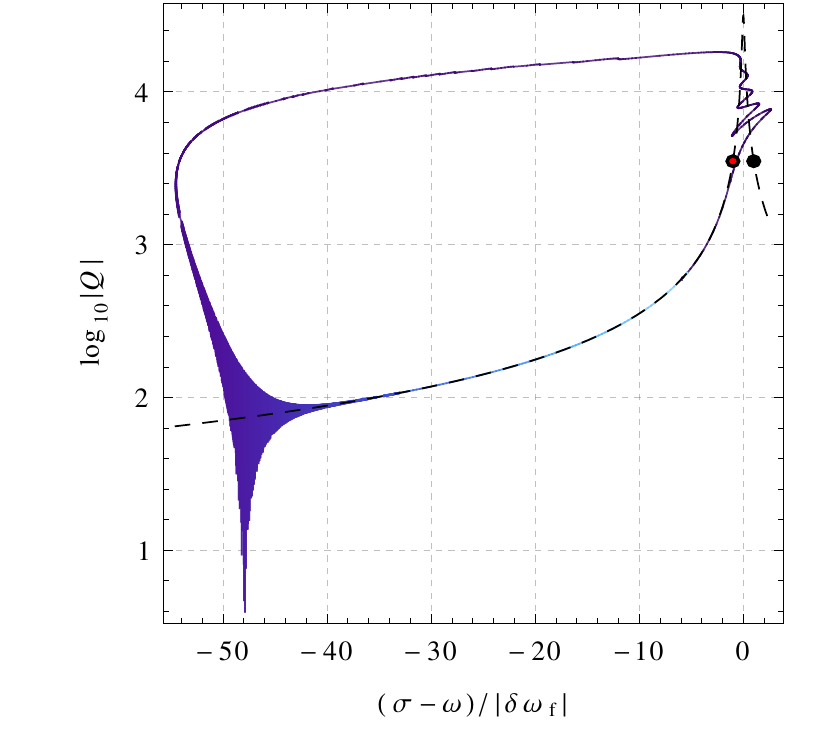}\\[.1cm]
  \includegraphics{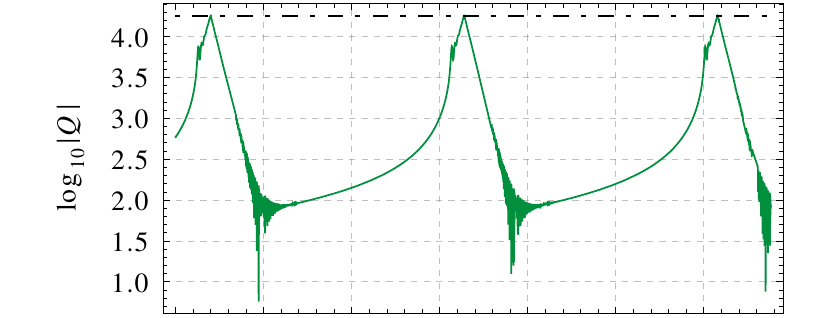}
  \includegraphics{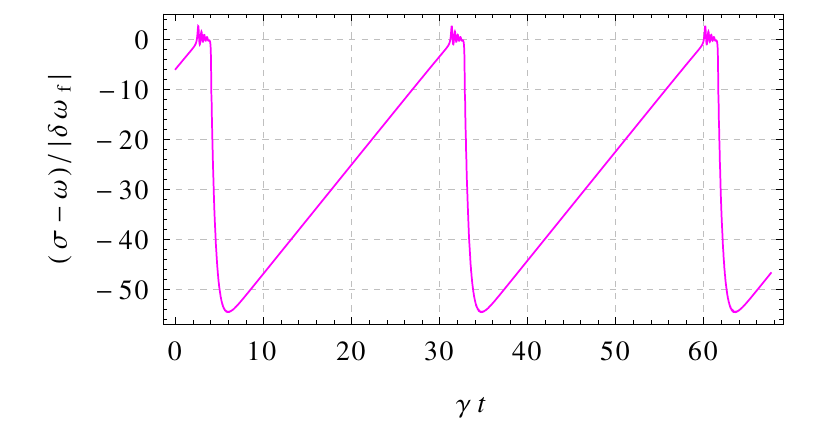}
  \caption{Resonance lock limit cycle for a case in which the fixed points exist (red and black circles) but are linearly unstable. Parameters used were $\Gdr/\omega=10^{-7.7}$, $\gamma/\omega=10^{-4.5}$, $\Gn/\omega=10^{-10}$, and $r=0.5$. Conventions used are the same as in \fig{f:trajplot1}. Color shows time, ranging from purple at $t=0$ to light blue. The red \& black circles show the lagging and leading fixed points, respectively (\se{s:rcap}). System sweeps through resonance without initially being captured. However, the system then oscillates through resonance several times, pumping up the mode's amplitude. Eventually, the oscillation ceases and the mode's angular momentum discharges into rotation causing the system to travel back away from resonance and start over. The lagging fixed point's eigenvalues are $\lambda/\omega \in \{-3.9\times10^{-4},\ (1.6\pm0.49\,i)\times10^{-4}\}$.}
  \label{f:trajplot2}
\end{figure}

\begin{figure}
  \includegraphics{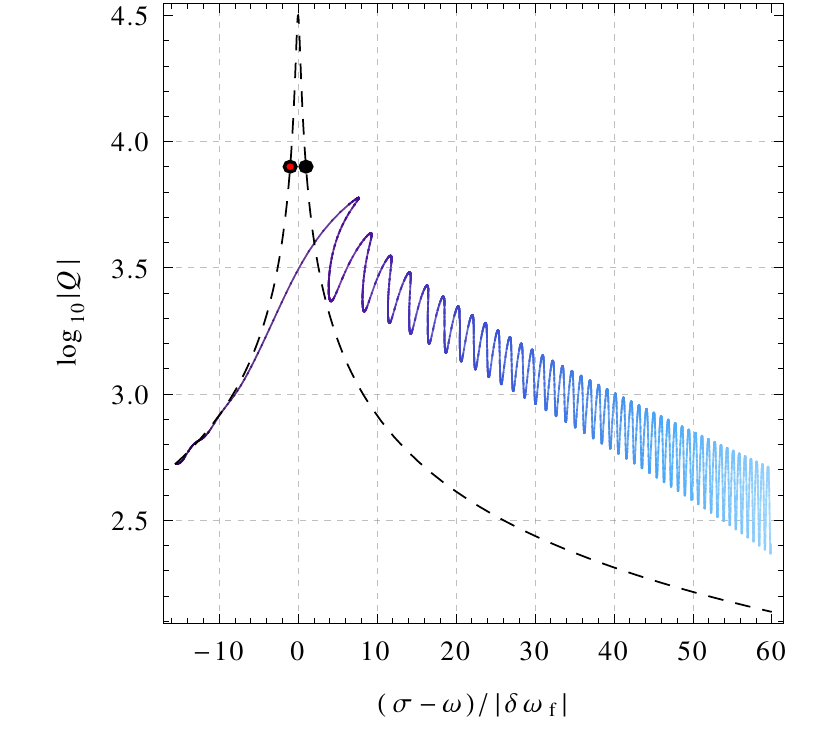}\\[.1cm]
  \includegraphics{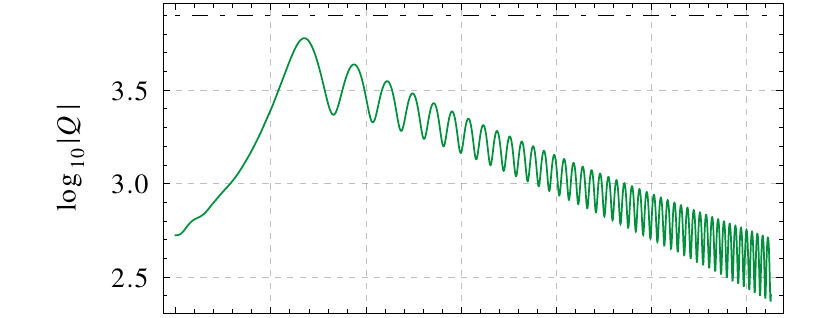}
  \includegraphics{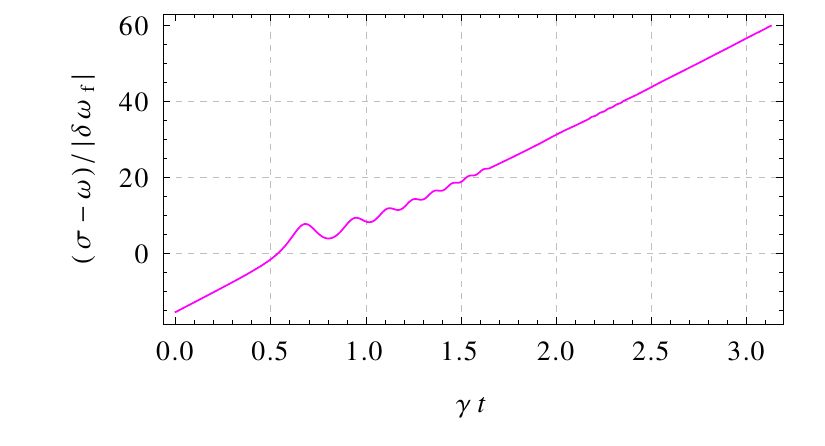}
  \caption{Failed resonance lock. Parameters used were $\Gdr/\omega=10^{-7}$, $\gamma/\omega=10^{-4.5}$, $\Gn/\omega=10^{-10}$, and $r=0.5$. Conventions used are the same as in \fig{f:trajplot1}. Color shows time, ranging from purple at $t=0$ to light blue. The unstable, repulsive fixed points (red and black circles) suppress the maximum attainable mode amplitude below the dash-dotted black line in the second panel, which shows the maximum amplitude that would be achieved without backreaction (\se{s:nbr}). This suppression is severe enough to prevent resonance locking from occurring. The lagging fixed point's eigenvalues are $\lambda/\omega \in \{-6.7\times10^{-4},\ 5.4\times10^{-4},\ 6.8\times10^{-5}\}$.}
  \label{f:trajplot3}
\end{figure}

\begin{figure}
  \includegraphics{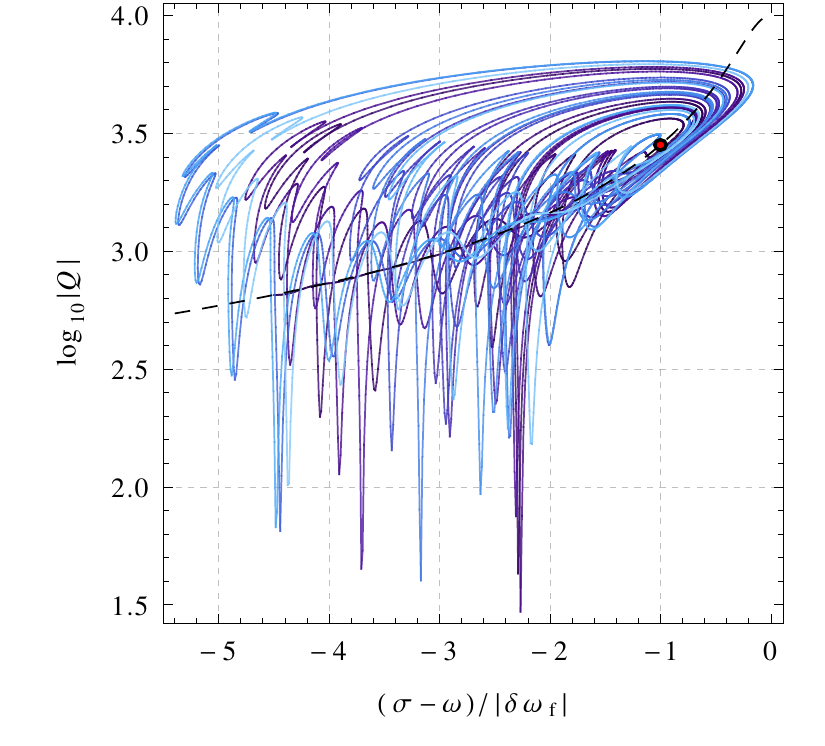}\\[.1cm]
  \includegraphics{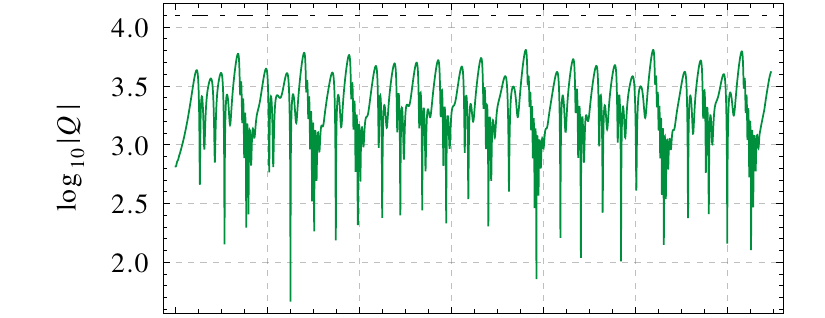}
  \includegraphics{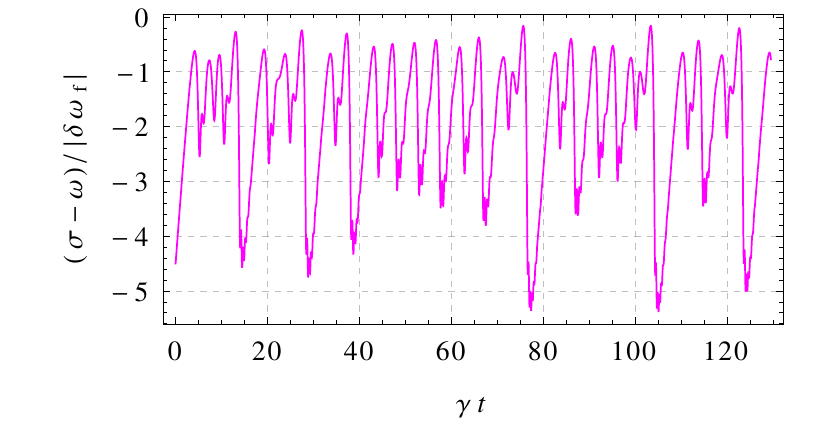}
  \caption{Chaotic resonance lock around an unstable fixed point (red circle). Parameters used were $\Gdr/\omega=10^{-7.4}$, $\gamma/\omega=10^{-4}$, $\Gn/\omega=10^{-10}$, and $r=0.5$. Conventions used are the same as in \fig{f:trajplot1}. Color shows time, ranging from purple at $t=0$ to light blue. This situation is similar to that depicted in \fig{f:trajplot2}, but in this case backreaction is so significant that the system never fully returns to the adiabatic approximation once deviating from it near the fixed point. Chaotic orbits are instead executed around the fixed point. The lagging fixed point's eigenvalues are $\lambda/\omega \in \{(0.98\pm2.4\,i)\times10^{-4},\ -4.0\times10^{-4}\}$.}
  \label{f:trajplot4}
\end{figure}

Figures \ref{f:trajplot1} -- \ref{f:trajplot4} show full numerical solutions to \eqs{e:eom2}{e:eomsig2} for several different choices of our four parameters $\gamma$, $\Gdr$, $\Gn$, and $r$. In each case, the mode amplitude is initialized to the adiabatic approximation in the regime where it should be valid (\se{s:sec}).

First, \fig{f:trajplot1} gives a simple example of resonance locking into a stable fixed point. Next, we hold $\Gn$, $\gamma$, and $r$ constant, but increase $\Gdr$ by a factor of $\sim10$, thus making $\sigma$ sweep towards resonance more quickly. In this case the fixed point is no longer stable, since \eq{e:estabul} is no longer satisfied, and \fig{f:trajplot2} shows the limit cycle resonance lock that then occurs. The system passes through resonance in between the two unstable fixed points, and then oscillates back and forth through resonance. This oscillation pumps the mode amplitude up as the system is repelled by the fixed points. Eventually, the oscillation ceases and the mode's angular momentum discharges into rotation causing the system to travel back away from resonance and decay back onto the adiabatic approximation. The cycle then begins again. This limit cycle is in fact precisely the stable periodic orbit generated by the supercritical Hopf bifurcation (\se{s:eig}).

\fig{f:trajplot3} shows the resulting evolution again holding all parameters constant except for $\Gdr$, which we increase by another factor of $\sim10$. The resonance locking fixed point is now sufficiently unstable that it suppresses the mode amplitude's growth and prevents resonance locking from occurring. Near resonance the mode amplitude still grows appreciably, but after the lock fails to hold, damping causes the mode amplitude to decay exponentially. We have found the linear character of the lagging fixed point to be the principal distinguishing factor between a limit cycle occurring and a failed resonance capture due to fixed point suppression. Specifically, we find that if the lagging fixed point is an unstable node, meaning all its eigenvalues are real (\fig{f:evals}), then it acts to suppress the mode amplitude and leads to a failed capture. If instead the lagging fixed point is an unstable spiral, meaning its eigenvalues contain an unstable complex conjugate pair (\se{s:eig}), then it is able to ``pump'' a trajectory to high amplitude and allow the formation of a limit cycle. This distinction, which is valid within a factor of $\sim3$ in $\Gdr$, will be critical in \se{s:achieveanalyt}.

\subsection{Chaos}\label{s:chaos}
Lastly, \fig{f:trajplot4} shows a chaotic trajectory. This situation is very similar to the limit cycle evolution shown in \fig{f:trajplot2}, in that the resonance locking fixed point is unstable but $\Gdr$ is not so large that resonance locking doesn't occur altogether; the essential difference is that the mode amplitude profile resulting from the adiabatic approximation (dashed line) is capped by damping not far above the fixed points, unlike in \fig{f:trajplot2} where it ascends much higher. This corresponds to the fact that the choice of parameters for \fig{f:trajplot4} lies close (logarithmically speaking) to the bifurcation manifold in parameter space where \eq{e:exist1} ceases to be satisfied and the fixed points no longer exist. This appears to be a key ingredient for chaos, as we will explain below, which is why we have changed $\gamma$ to a larger value than that used in Figures \ref{f:trajplot1} -- \ref{f:trajplot3} (so that \eqp{e:exist1} is closer to not being satisfied).

Since the fixed points are so close to the peak of the adiabatic profile, the pumping process that occurs due to repulsion from the fixed points cannot allow the mode to acquire a very large amplitude. As a result, when the mode's angular momentum eventually drains into the background spin, the system does not retreat back from resonance very far, and has little time to decay back onto the adiabatic approximation before resonance is reached again. The initial condition upon entering resonance is consequently somewhat different each cycle, leading to the potential for chaos.

\begin{figure}\centering
  \begin{overpic}[width=237.6pt,height=182.8pt]{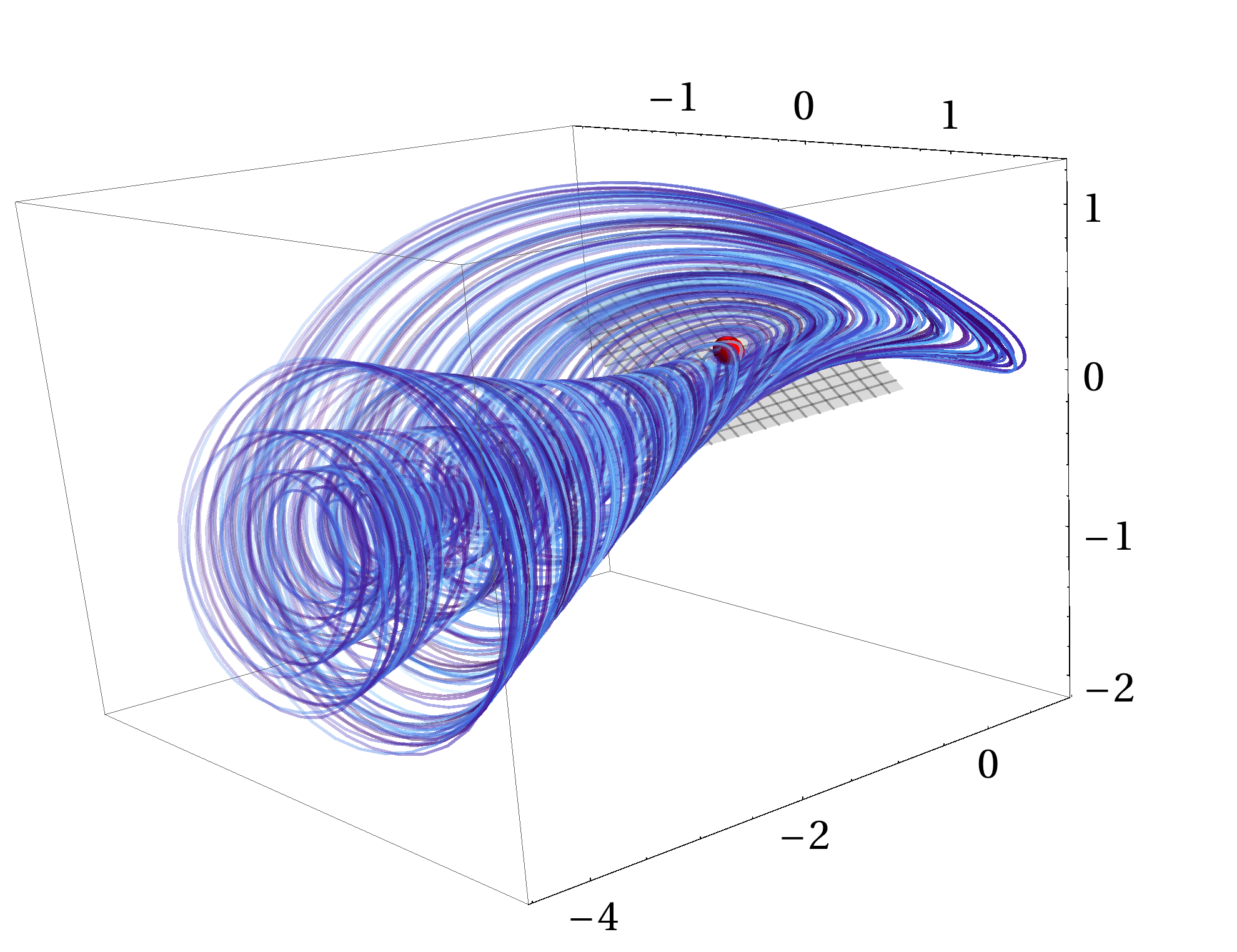}
   \put(62,3){\rotatebox{20}{$(\sigma-\omega)/|\domegf|$}}
   \put(57,73){\rotatebox{-4}{$\Im(Q-Q_\f)/|Q_\f|$}}
   \put(94,50){\rotatebox{-90}{$\Re(Q-Q_\f)/|Q_\f|$}}
  \end{overpic}
  \caption{Three-dimensional projection of the integration from \fig{f:trajplot4}. Time is indicated by color, ranging from dark purple at early times to light blue. The unstable plane corresponding to eigenvalues $(0.98\pm2.4\,i)\times10^{-4}$ is shown, centered on the fixed point (red sphere). Each cycle, the system begins near the fixed point, but is then ejected along the unstable manifold. Nonlinear terms cause the system to decay back onto the adiabatic solution; this motion comprises the spiral structure on the left. The adiabatic solution then transports the system near to the fixed point, and the cycle begins again with perturbed initial conditions.}
  \label{f:chaosplot}
\end{figure}

We show a three-dimensional projection of the orbit from \fig{f:trajplot4} in \fig{f:chaosplot}. The lagging fixed point is shown by a small red sphere, while its unstable plane corresponding to eigenvalues $\lambda/\omega=(0.98\pm2.4\,i) \times 10^{-4}$ is also displayed.

We now present numerical evidence that the path depicted in \fig{f:chaosplot} follows a strange attractor. We emphasize that our evidence is not rigorous. A strange attractor of a dynamical system, also known as an attracting chaotic invariant set, is a set that \citep{wiggins}:
\begin{enumerate}
 \item is compact,
 \item is invariant under the dynamical equations,
 \item is attracting,
 \item has sensitive dependence on initial conditions, and
 \item is topologically transitive.
\end{enumerate}

\begin{figure}\centering
  \begin{overpic}{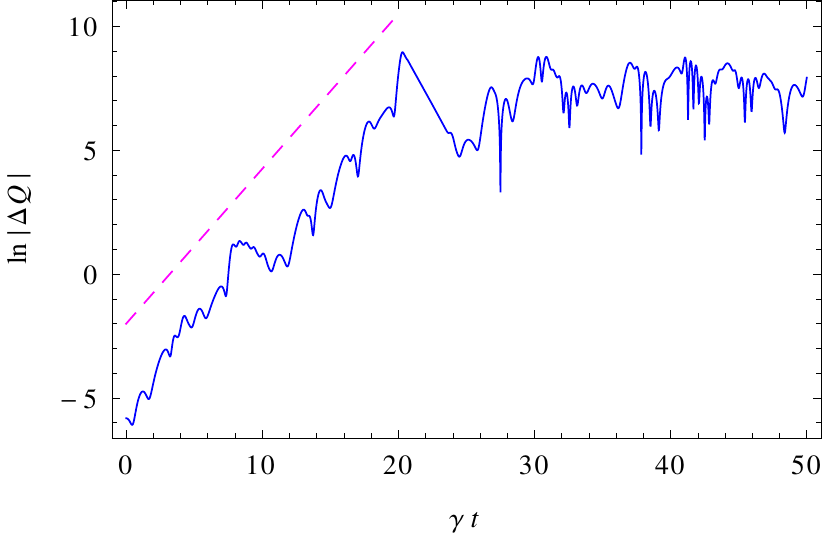}
  \put(25,18){\rotatebox{49}{Exponential divergence}}
  \end{overpic}
  \caption{Estimation of largest Lyapunov exponent for the chaotic trajectory in Figures \ref{f:trajplot4} \& \ref{f:chaosplot}. We initialized two numerical integrations of our dynamical equations on the adiabatic solution (\se{s:sec}) with slightly perturbed initial detuning frequencies: $\domeg_0 = 0.95\domegf$ and $0.950001\domegf$. The blue curve shows the norm of the difference between the resulting values of the reduced mode amplitude $Q$ as a function of time. The trajectories initially deviate exponentially, demonstrating chaos, with a rough functional form of $e^{t/\tau}$ for $\tau\approx2\times10^4/\omega$ (dashed magenta line). We thus estimate the largest Lyapunov exponent \citep{wiggins} for the trajectories to be $\approx1/\tau\approx6\times10^{-5}\omega>0$, which is close to the damping rate of $\gamma=10^{-4}\omega$.}
  \label{f:lyapunov}
\end{figure}

\fig{f:chaosplot} appears to begin to trace out a bounded, attracting, invariant set, which we denote $\Lambda$; this addresses conditions (i) -- (iii). In \fig{f:lyapunov}, we estimate the largest Lyapunov exponent of the trajectory in \fig{f:chaosplot} to be $6\times10^{-5}\omega$. Since this is positive, trajectories that begin together deviate exponentially as time passes. This then points toward condition (iv) being satisfied. Lastly, since color indicates time in \fig{f:chaosplot}, the fact that dark purple (early times) is tightly and randomly interwound with light blue (late times) leads us to believe that condition (v) is likely also satisfied. Again, we have presented only suggestive evidence; further study is required to fully address the presence of a strange attractor.

In addition, we note that our preliminary investigations show that the chaos present results from the Hopf orbit undergoing a period-doubling cascade, and is similar in several ways to the R\"ossler attractor \citep{rossler76}. This requires further study.

\section{Achieving resonance locks}\label{s:achieve}
\subsection{Numerical results}\label{s:achievenum}
In order to ascertain the conditions that lead to resonance locks, we performed sample integrations of \eqs{e:eom2}{e:eomsig2} numerically. We initialized each integration with
\begin{displaymath}
 \domeg_0 = 10\times\max\left(\domegf,\domegad, \gamma  \right)
\end{displaymath}
and $\Q_0$ set by the adiabatic approximation from \eq{e:sec}. We then performed each integration until $t_1=2\domeg_0/\omega\Gdr$. We determined that a resonance lock occurred if, assuming $r<1$,
\begin{displaymath}
 \frac{\min\domeg}{\domeg_0} > -0.5,
\end{displaymath}
where $\min\domeg$ is the minimum value of $\domeg$ attained over the final 10\% of integration.\footnote{Numerous other potential resonance lock criteria exist; however, we have found this to be the most reliable.} A similar formula was used for $r>1$, but accounting for the fact that resonance is then approached from the left (in terms of $\domeg$; see \se{s:exist}). Note that these conditions account for stable, limit cycle, and chaotic forms of resonance locking (\se{s:trajnum}).

\begin{figure*}
\centering
  \vspace{.3cm}
  \rotatebox{90}{\rule{65pt}{0pt}$\log_{10}\left(\Gdr/\omega\right)$}
  \begin{overpic}{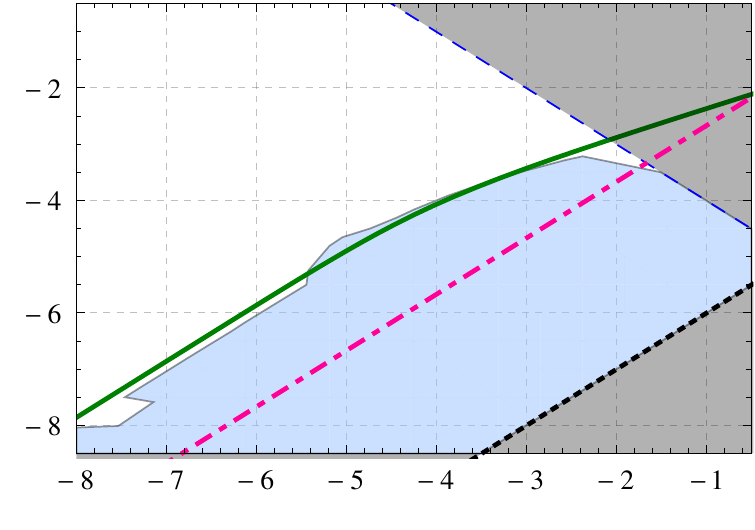}
   \put(27,52){No locks}
   \put(60,20){\rotatebox{32}{Stable locks}}
   \put(31,20){\rotatebox{32}{Limit cycle/chaotic locks}}
   \put(73,60){No fixed point}
   \put(13,72){$\Gn/\omega=10^{-5}$, $r=0.01$}
   \put(45,0){$\log_{10}(\gamma/\omega)$}
  \end{overpic}\rule{5pt}{0pt}
  \rotatebox{90}{\rule{65pt}{0pt}$\log_{10}\left(\Gdr/\omega\right)$}
  \begin{overpic}{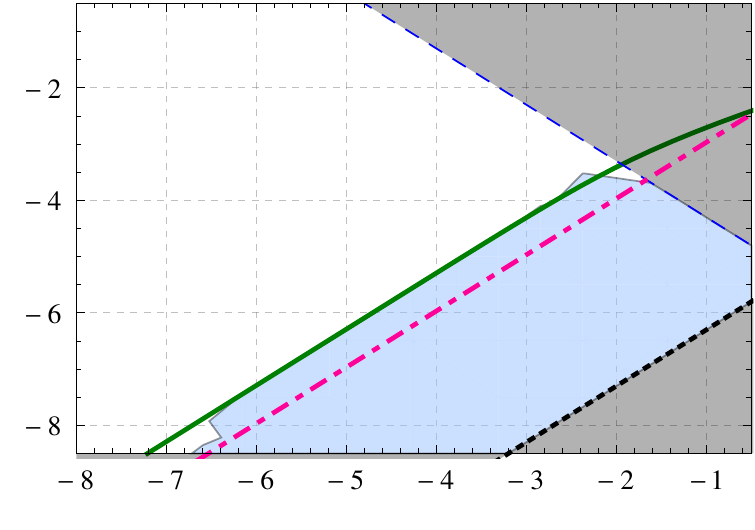}
   \put(13,72){$\Gn/\omega=10^{-5}$, $r=0.5$}
   \put(45,0){$\log_{10}(\gamma/\omega)$}
  \end{overpic}\\[17pt]
  \rotatebox{90}{\rule{65pt}{0pt}$\log_{10}\left(\Gdr/\omega\right)$}
  \begin{overpic}{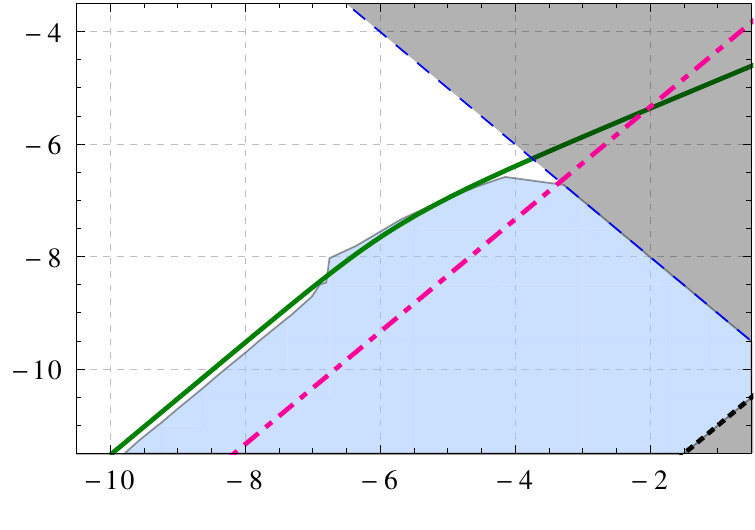}
   \put(13,72){$\Gn/\omega=10^{-10}$, $r=0.01$}
   \put(45,0){$\log_{10}(\gamma/\omega)$}
  \end{overpic}\rule{5pt}{0pt}
  \rotatebox{90}{\rule{65pt}{0pt}$\log_{10}\left(\Gdr/\omega\right)$}
  \begin{overpic}{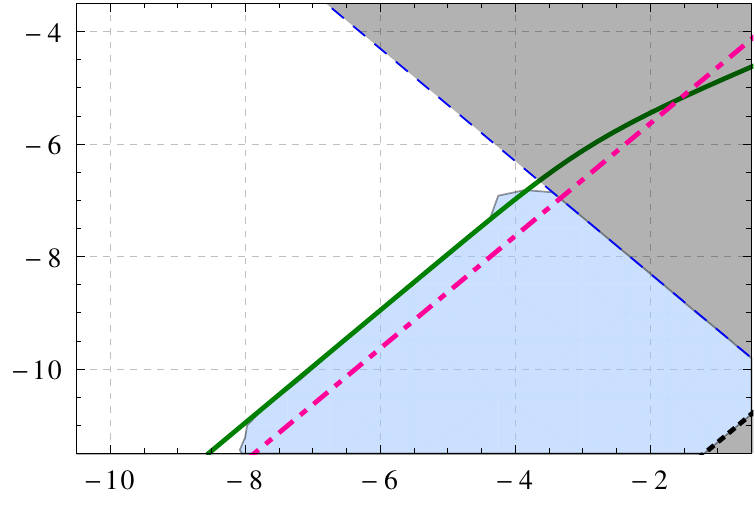}
   \put(13,72){$\Gn/\omega=10^{-10}$, $r=0.5$}
   \put(45,0){$\log_{10}(\gamma/\omega)$}
  \end{overpic}\\[17pt]
  \rotatebox{90}{\rule{65pt}{0pt}$\log_{10}\left(-\Gdr/\omega\right)$}
  \begin{overpic}{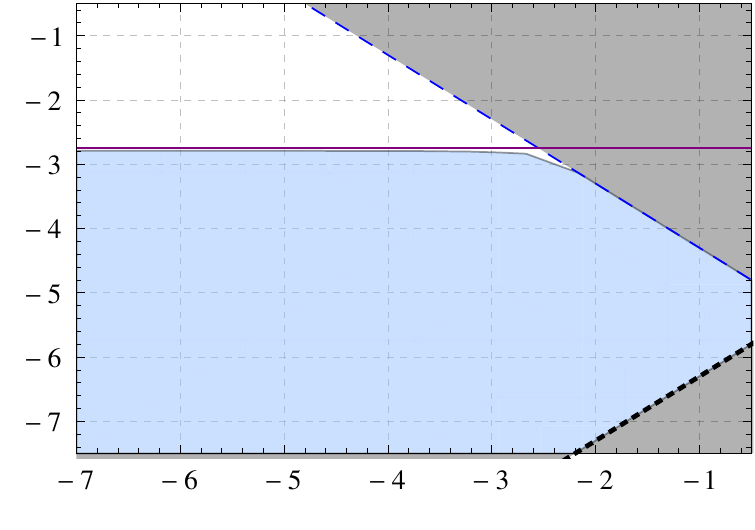}
   \put(13,72){$\Gn/\omega=10^{-5}$, $r=1.5$}
   \put(45,0){$\log_{10}(\gamma/\omega)$}
   \put(40,28){Stable locks}
  \end{overpic}\rule{5pt}{0pt}
  \rotatebox{90}{\rule{65pt}{0pt}$\log_{10}\left(-\Gdr/\omega\right)$}
  \begin{overpic}{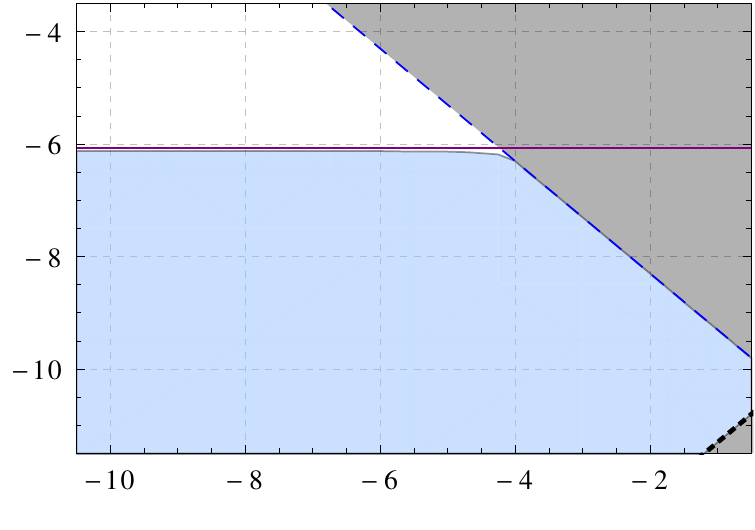}
   \put(13,72){$\Gn/\omega=10^{-10}$, $r=1.5$}
   \put(45,0){$\log_{10}(\gamma/\omega)$}
  \end{overpic}
  \caption{Resonance lock regimes based on numerical solution of mode, spin, and orbit evolution equations from \se{s:exist}. Light blue shading indicates that resonance locking occurs (including stable fixed point locks, limit cycles, and chaotic locks, as in Figures \ref{f:trajplot1}, \ref{f:trajplot2}, \& \ref{f:trajplot4}), while white indicates the reverse (as in \fig{f:trajplot3}). Above each dashed blue line, the resonance locking fixed point does not exist (\se{s:exist}; \eqp{e:exist1}). The green and purple lines correspond to our analytic formulae for the resonance locking regime (\se{s:achieveanalyt}; equations \ref{e:capsucc} \& \ref{e:capsuccret} respectively). The dash-dotted magenta line indicates the upper boundary of the domain of stability of the resonance lock fixed point (\se{s:stab}; \eqp{e:estabul}), while the dotted black line shows where the weak-tide limit is certainly violated (\se{s:exist}; \eqp{e:weaktide}). Limit cycles and chaotic orbits (as in Figures \ref{f:trajplot2} \& \ref{f:trajplot4}) preferentially occur in the regions between the green and magenta lines (where the fixed point is unstable but locks still occur).}
  \label{f:capregime}
\end{figure*}
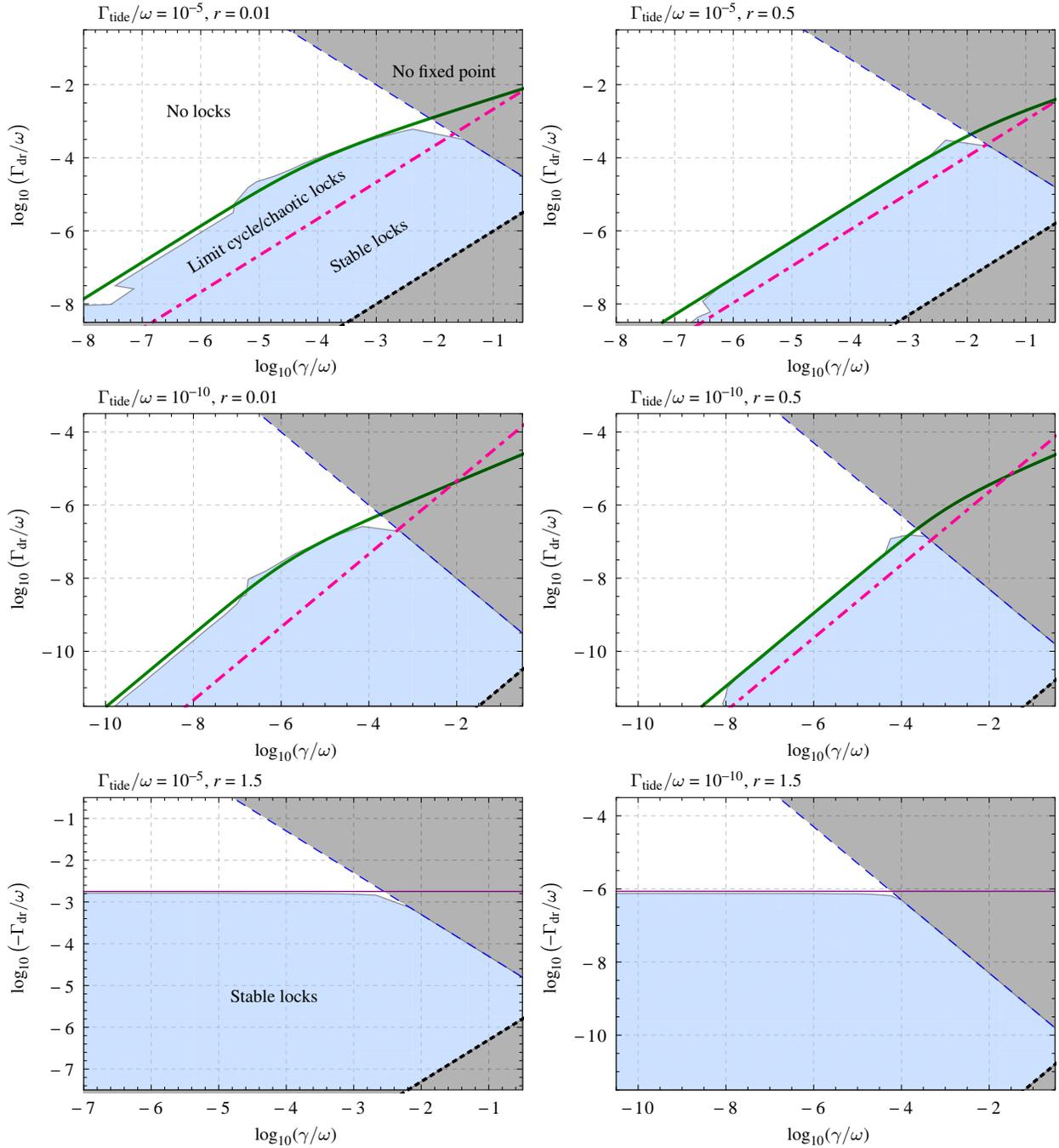

\fig{f:capregime} shows our results. Light blue regions indicate that a resonance lock did occur, while white indicates the reverse, and dark gray indicates that the fixed point does not exist (\se{s:exist}). The green lines are the analytic formula from \eq{e:capsucc} below for the boundary between successful and failed resonance locking. The thin purple lines are the equivalent condition for resonance locks to occur when $r>1$, from \eq{e:capsuccret}. Both analytic approximations, which we will develop in the next section, show good agreement with numerical results.

\subsection{Analytic approximations}\label{s:achieveanalyt}
We now seek to obtain an analytical understanding of our numerical results in \fig{f:capregime}. We will thus attempt to assemble a set of analytic approximations to determine what values of our four parameters $\gamma$, $\Gdr$, $\Gn$, and $r$ (\se{s:form}) lead to resonance locks, and what values do not. We define resonance locking in this context to be any behavior such that $\sigma$ does not increase without bound as $t\rightarrow\infty$; this definition comprises locking into a stable fixed point (\fig{f:trajplot1}), limit cycles about an unstable fixed point (\fig{f:trajplot2}), and chaotic behavior like in \fig{f:trajplot4}, but does not include the behavior seen e.g.\ in \fig{f:trajplot3}.

We see by inspecting \eqs{e:Qf}{e:sec} that the adiabatic solution exactly passes through the resonance locking fixed point. As a result, a sufficient condition for a resonance lock to be achieved is $|\domegf|>|\domeg_\ad|$, which evaluates to
\begin{equation}
 \Gdr^2 \lesssim (1-r)|\Gn| \gamma,
\end{equation}
together with fixed point stability (\se{s:stab}; \eqp{e:estabul}). However, this is a very conservative estimate of the resonance locking regime. To develop a set of necessary and sufficient criteria for resonance locks, recall that the time derivative of $\domeg$ is given by (\eqp{e:eomsig2})
\begin{equation}\label{e:eomsig3}
 \dot{\domeg}/\omega = -\Gdr + \Gn(g_1-rg_2),
\end{equation}
where $g_1$ and $g_2$ are
\begin{equation}
 g_1 = \frac{\gamma}{\omega}|\Q|^2 \qquad g_2=\Im[\Q].
\end{equation}

First, assume $r\gg1$. By \eq{e:exist2} we see that $\Gdr<0$ and thus that the system approaches resonance from the left (in terms of $\domeg$; note that in Figures \ref{f:trajplot1} -- \ref{f:trajplot4} the abscissa is $\sigma-\omega=-\domeg$). Since the lagging fixed point is always stable in this situation (\se{s:eig}), we simply need the resonance passage to provide sufficient amplitude to reach the fixed point in order for a stable resonance lock to take hold. Invoking the no-backreaction approximation results from \se{s:nbr} to substitute a maximum value of $\Im[\Q]$ into \eq{e:eomsig3}, dropping $g_1$, and setting $\dot{\domeg}\sim0$ leads to the following condition for resonance locking:
\begin{equation}\label{e:capsuccret}
 -\frac{\Gdr}{\Gn} > 3r\sqrt{-\frac{\pi\omega}{\Gdr}}.\qquad (r\gg1)
\end{equation}
Although our analysis is strictly valid only for $r$ much larger than unity, we find it to work well even for $r\gtrsim1$, as can be seen in the bottom row of \fig{f:capregime} (where $r=1.5$). We have inserted an additional factor of $3$ on the right-hand side of \eq{e:capsuccret} to match our numerical results.

Next, assume $r\ll1$. By \eq{e:exist2} this means $\Gdr>0$, so the system approaches the resonance locking fixed point from the right (in terms of $\domeg$). Here, however, the fixed point is \emph{not} always stable, as we found in \se{s:eig}. In \se{s:trajnum}, we argued that resonance locks in the form of limit cycles or chaotic trajectories could occur when the fixed point was an unstable spiral (with a complex conjugate pair of eigenvalues), but that resonance capture failed when the fixed point was an unstable node (with all real eigenvalues); this was true within a factor of $\sim3$ in terms of the value of $\Gdr$. Thus a necessary condition for resonance locks (to within a factor of $\sim3$) is \eq{e:ccsplit}, which specifies where the fixed point is a spiral. Next, we drop $g_2$ and again substitute the no-backreaction approximation results from \se{s:nbr} to find that resonance passage can deliver a system to the resonance locking fixed point if
\begin{displaymath}
\frac{\gamma}{\Gdr} > \frac{\Gdr}{2\pi\Gn}.
\end{displaymath}
Augmented with \eq{e:ccsplit}, this approximately becomes the following condition for resonance locking:
\begin{equation}\label{e:capsucc}
 \frac{\gamma}{\Gdr} > \frac{\Gdr}{6\pi\Gn} + \frac{1}{3}\left(\frac{r^{2/3}}{1-r}\right)\left( \frac{\omega}{\Gn} \right)^{1/3}\!\!\!.\qquad(r\ll1)
\end{equation}
Similar to our formula for $r>1$, our analysis is valid only for $r$ very close to zero; however, we again find it to work well even for $r\lesssim1$, as can be seen in the right panels of the top two rows of \fig{f:capregime} (where $r=0.5$). We have inserted an additional factor of $1/3$ on the right-hand side of \eq{e:capsucc} to match our numerical results.

\section{Tidal evolution during resonance locks}\label{s:evolve}
\subsection{Accelerating tidal evolution}\label{s:amp}
Here we generalize the energetic arguments made in \citet{burkart13} to estimate the orbital and spin evolution during a resonance lock. During a lock, the reduced mode amplitude $Q$ is roughly given by its value at the fixed point, i.e., \eq{e:sol2}; note, however, that this approximation is very crude for limit cycles and chaotic orbits (e.g.\ Figures \ref{f:trajplot2} \& \ref{f:trajplot4}) since in those cases the real and imaginary parts of $Q$ have a more complicated dependence on time. Using \eqs{e:Omegdotorb}{e:sol2} together with our definitions of $\Gdr$, $\Gn$, and $r$ from equations \eqref{e:Gdr} -- \eqref{e:r}, we can derive
\begin{equation}\label{e:Omegdotorbrl}
 \dot{\Omega}_\orb = \left( \frac{1}{1-r} \right)\left[ \alpha_\orb - \frac{r}{k}\left( m \alpha_\spin+\frac{\partial \omega}{\partial t}\right)\right],
\end{equation}
where again $\alpha_\orb$ and $\alpha_\spin$ represent the contributions to $\dot{\Omega}_\orb$ and $\dot{\Omega}_\spin$ from slowly varying processes other than resonant interaction with the normal mode in question (see \eqsp{e:Omegdotspin}{e:Omegdotorb}). This can be converted into an energy transfer rate by \eq{e:EOmeg}. Performing a similar derivation for the spin frequency, we have
\begin{equation}\label{e:Omegdotspinrl}
 \dot{\Omega}_\spin = \left( \frac{r}{1-r} \right)\left[ - \alpha_\spin + \frac{1}{rm}\left( k \alpha_\orb-\frac{\partial \omega}{\partial t}\right)\right],
\end{equation}

Examining \eqs{e:Omegdotorbrl}{e:Omegdotspinrl}, we see that a resonance lock acts to \emph{accelerate} the orbital and spin evolution given by the nonresonant processes contributing to $\alpha_\orb$ and $\alpha_\spin$, which are due e.g.\ to gravitational wave orbital decay, the equilibrium tide, etc. Moreover, since the time derivative of the eccentricity is simply a linear combination of $\dot{\Omega}_\orb$ and $\dot{\Omega}_\spin$ \citep{witte99}, resonance locking also accelerates circularization. The degree of acceleration depends on how close the moment of inertia ratio $r$ is to unity; we estimate under what conditions $r\sim1$ in \se{s:rval}. This acceleration of tidal evolution is what led \citet{witte02} to conclude that resonance locks solve the solar-binary problem \citep{meibom06}, although they neglected essential nonlinear effects that obviate their results (see \se{s:assump}).

The presence of $\alpha_\spin$ in the evolution equation for $\dot{\Omega}_\orb$ (and $\alpha_\orb$ in the equation for $\dot{\Omega}_\spin$) implies that a resonance lock efficiently couples orbital and spin evolution together, as well as to stellar evolution through the rate of change of the eigenfrequency $\partial\omega/\partial t$. For example, if gravitational waves in an inspiraling white dwarf binary cause orbital decay, a resonance lock will cause tidal synchronization to occur on a gravitational wave timescale (\se{s:wd}). Similarly, resonance locking can cause stellar spindown by magnetic braking or eigenfrequency evolution due to tidal heating to backreact on the binary orbit.

\subsection{Conditions for rapid tidal evolution}\label{s:rval}
Here we estimate the ``typical'' value of $r$ (defined in \eqp{e:r}) expected to occur in a given binary, so as to assess whether resonance locks significantly accelerate tidal evolution (\se{s:amp}). First, consider a binary in a circular, spin-aligned orbit. For the lowest-order $l=2$ spherical harmonic of the tidal potential, only one temporal Fourier component of the tidal forcing exists: $k=m=2$ (\se{s:modeamp}). With only a single forcing component, it is thus generically unlikely to find $r\sim1$. The exception is when considering a star whose companion's mass is $\ll M$, in which case $r$ may be $\sim1$ even for a circular orbit.

For an eccentric orbit, however, there is significant power at many harmonics $k$ of the orbital frequency. In this case, a resonance lock persists until it is disrupted when another mode, driven by a different Fourier component, also comes into resonance and upsets the balance of spin and orbital evolution theretofore enforcing $\dot{\domeg}=0$. Whether the existing resonance lock can withstand a second resonance passage depends on how ``robust'' it is; we can estimate the degree of ``robustness'' using the maximum amplitude achieved under the no-backreaction approximation from \se{s:nbr}:\footnote{Our results are insensitive to the expression by which a resonance lock's ``robustness'' is quantified, so long as it is proportional to $U$.}
\begin{equation}\label{e:S1}
   |q|_\mr{max} = |U|\sqrt{\frac{2\pi\omega}{|\Gdr|}}.
\end{equation}

If we equate $\omega\approx k\Omega_\orb$, having taken $|m\Omega_\spin|\ll |k\Omega_\orb|$ for simplicity, then $|q|_\mr{max}$ depends on the harmonic index $k$ as
\begin{equation}\label{e:S2}
 |q|_\mr{max} \propsim k^b X_{lm}^k,
\end{equation}
where $X$ is a Hansen coefficient and $b>0$ is a constant that accounts for power law dependences on the tidal overlap integral and other mode-dependent quantities entering into \eq{e:S1} (\se{s:modeamp}). Using our scaling derived in \app{a:hansen}, \eq{e:S2} becomes
\begin{equation}
 |q|_\mr{max}\propsim k^{b-1/2} \exp\left[ -k g(e) \right],
\end{equation}
where the full form of $g(e)\approx (1-e^2)^{3/2}/3$ is given in \eq{e:g}.

The longest-lived resonance locks in an eccentric binary will be those with the largest values of $|q|_\mr{max}$. Thus we can estimate the ``typical'' value of $r$ by finding the value of $k$ that maximizes $|q|_\mr{max}$. This is
\begin{equation}
\mr{argmax}_k |q|_\mr{max} \approx \frac{b-1/2}{g(e)}.
\end{equation}
In order to produce a simple order-of-magnitude estimate, we take $b-1/2\sim1$ to find that the most robust resonance locks will have  $r\sim1$, and thus significantly increase the rate of tidal evolution, if\footnote{Note that whether $r>1$ or $r<1$ is determined by the sign of $\Gdr$, which is independent of our estimates here; see \se{s:exist}.}
\begin{equation}
 1-e^2 \sim \left(10\frac{\Is}{\mu a^2}\right)^{1/3}.
\end{equation}
When this condition is not satisfied, then either the longest-lived resonance locks will not provide much acceleration of tidal evolution, or there will be no long-lived locks at all.

\section{Astrophysical applications}\label{s:astroph}
In \S\se{s:wd} \& \ref{e:ecc}, we determine where tidal resonance locks may be able to occur by applying the criteria we have developed---\eqs{e:exist1}{e:capsucc}---to inspiraling compact object binaries and eccentric stellar binaries.

\subsection{Inspiraling compact object binaries}\label{s:wd}
\begin{figure}
  \begin{overpic}{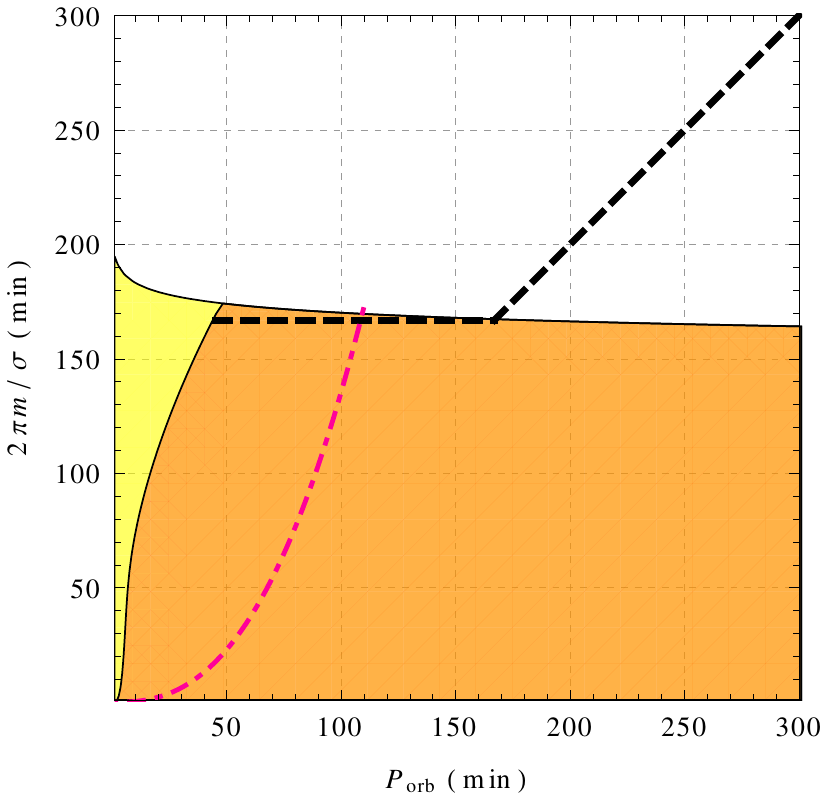}
    \put(58,32){Stable locking}
    \put(19,18){\rotatebox{68}{Limit cycle/chaotic locking}}
  \end{overpic}\\[.2cm]
  \caption{Resonance locking regimes for a fiducial white dwarf binary. Yellow shading indicates that the resonance locking fixed points exists (\eqp{e:exist1}), but that resonance locks are nonetheless not possible. Orange shading indicates the complete resonance locking regime, derived from our numerical results in \se{s:achievenum} (\eqp{e:capsucc}). The black dashed line shows a schematic system trajectory, assuming the initial spin is negligible. The dot-dashed magenta line shows where the lagging fixed point becomes unstable (\eqp{e:estabul}).}
  \label{f:wdns}
\end{figure}

We consider the case of a circular, spin-orbit-aligned binary consisting of either two white dwarfs or two neutron stars. The equivalent of the drift rate $\Gdr$ in this case is
\begin{equation}
 \Gdr=\frac{m\Omega_\orb}{\omega\tgw},
\end{equation}
where $\tgw$ is the gravitational wave orbital decay time given by \citep{peters64}
\begin{equation}
\tgw = \omegs^{-1}\frac{5}{96}\frac{\left( 1+M'/M \right)^{1/3}}{M'/M}\betas^{-5}\left( \frac{\omegs}{\Omega_\orb} \right)^{8/3}.
\end{equation}
Here $\omegs^2=GM/R^3$ is the dynamical frequency and $\betas^2=GM/Rc^2$ is the ratio of the escape velocity to the speed of light. We consider only the $l=|m|=2$ component of the tidal response.

For a fiducial double-white dwarf binary, we use the $0.6 M_\odot$, $T_\mr{eff}=5,500$~K carbon/oxygen white dwarf model described in \citet{burkart13}, with a radius of $R=0.013R_\odot$, and a moment of inertia of $I_*=0.18 MR^2$. For the damping rate $\gamma$ and tidal overlap integral $\Qol$, we use the following scaling relations listed in Table 3 of \citet{burkart13}:
\begin{align*}
 \Qol&\sim 27\times\left(\frac{\sigma}{\omega_*}\right)^{3.69}\\
 \gamma &\sim 2.9\times10^{-14}\omega_*\times\left(\frac{\sigma}{\omega_*}\right)^{-1.88},
\end{align*}
where our mode normalization convention is $\Es=\epsilon$ and we are neglecting rotational modifications of the stellar eigenmodes (in other words, setting the Coriolis force operator $B$ from \se{s:modeamp} to zero).

For a fiducial double-neutron star binary, we use the $M=1.4M_\odot$, $R=12$ km cold neutron star model employed in \citet{weinberg13}, which assumed the Skyrme-Lyon equation of state \citep{chabanat98,steiner09}. We assume that $I_*=0.18 MR^2$, as with our white dwarf model. \citet{weinberg13} give the following scaling relations for $l=2$ g-modes:
\begin{align*}
 \Qol&\sim 0.3\times\left(\frac{\sigma}{\omega_*}\right)^{2}\\
 \gamma &\sim 4\times10^{-8}\times\left(\frac{\sigma}{\omega_*}\right)^{-2}T_8^{-2}\ \mr{Hz},
\end{align*}
where $T_8$ is the core temperature in units of $10^8$~K.

\fig{f:wdns} shows our result for the white dwarf case. The yellow region shows where the resonance locking fixed point exists (\eqp{e:exist1}) but where resonance locks are nonetheless not possible according to our numerical and analytic results in \se{s:achieve}. The orange region shows the complete resonance locking regime derived from our numerical results (\eqp{e:capsucc}). The dashed black line in the top panel is a simple, schematic system trajectory for a double-white dwarf binary. The system begins in the upper right with a long orbital period and a small rotation rate. Once orbital decay causes the system to reach the orange region, a resonance lock occurs and the forcing frequency $\sigma$ is held approximately constant. This was already demonstrated in \citet{burkart13} using the adiabatic approximation (\se{s:sec}), which is valid for determining when the resonance locking fixed points exist.

Eventually, however, the system exits the orange region. At this point resonance locking is no longer possible because of the short gravitational wave inspiral time. This novel prediction comes from our analysis in this work.\footnote{\citet{burkart13} did assess where the adiabatic approximation (which was inaccurately referred to as the ``secular approximation'') became invalid; however, this is an incomplete consideration that fails to account for fixed point instability, limit cycles, etc.} Note that this prediction neglects nonlinear hydrodynamical phenomena (\se{s:assump}): in reality the wave amplitude eventually becomes large enough to cause wave breaking, as shown in \citet{burkart13,fuller12}. This may be a more stringent constraint on the existence of resonance locks in many close white dwarf binaries.

In the neutron star case, the gravitational wave time is much shorter than for white dwarf binaries at comparable values of $R/a$, since neutron stars are much more relativistic objects. Resonance passage thus happens very quickly, which prevents modes from reaching amplitudes large enough to allow locking. As a result, resonance locks are never possible, even though the resonance locking fixed point exists for $P_\orb\lesssim50$~ms.

To get a sense of the degree to which \eq{e:capsucc} fails to be satisfied for neutron star binaries, we first compute the following quantities with $\Omega_\spin=0$:
\begin{align*}
 \frac{\gamma}{\omega} &= 10^{-7}T_8^{-2} \left( \frac{P_\orb}{50\ \mr{ms}} \right)^3 & \frac{\Gdr}{\omega}&=6\times10^{-5}\left( \frac{P_\orb}{50\ \mr{ms}}
 \right)^{-5/3}\\
 \frac{\Gn}{\omega}& = 10^{-11}\left( \frac{P_\orb}{50\ \mr{ms}} \right)^{-6} & r &= 0.002\left( \frac{P_\orb}{50\ \mr{ms}} \right)^{-4/3}.
\end{align*}
We thus set $r\approx0$. Substituting the remaining values into \eq{e:capsucc} and simplifying, we have
\begin{equation}
 10\,T_8^2 < 10^{-7}\left( \frac{P_\orb}{50\ \mr{ms}} \right)^{1/3}.
\end{equation}
This shows that resonance locking fails to occur by $\sim8$ orders of magnitude at a wide range of orbital periods. The conclusion that resonance locks cannot occur in neutron star binaries is thus very robust.

\subsection{Eccentric binaries}\label{e:ecc}
In this section we estimate whether resonance locking can occur in eccentric stellar binaries \citep{witte99}. For a fiducial system, we take parameters consistent with KOI-54 \citep{welsh11}, where it has been recently suggested that one or more of the observed tidally excited pulsations may be the signatures of resonance locks (\citealt{fullerkoi, burkart12}; see however \citealt{oleary13}). This system consists of two similar A stars with $M\approx2.3M_\odot$, $R\approx2.2R_\odot$, and $T_\mr{eff}\approx8,500$~K. The binary's orbital parameters are $e=0.83$ and $P_\orb=43$~days. \citet{burkart12} estimated damping rates for such stars to be
\begin{equation}
 \gamma \sim 0.1 \left( \frac{\omega}{\omega_*} \right)^{-4}\ \mr{Myr}^{-1}.
\end{equation}
For overlap integrals $\Qol$, we take the following scaling derived for A stars from \citet{burkart12}:
\begin{equation}
   \Qol\sim10^{-5}\left( \frac{\omega}{\omega_*} \right)^{11/6}.
\end{equation}

The drift rate $\Gdr$ in this case comes from the equilibrium tide's influence on each star's spin and on the overall orbital frequency. We account only for the equilibrium tide's effect on the orbital frequency for simplicity. Parameterizing the equilibrium tide's energy transfer rate by its quality factor $\Qeq$, so that \citep{goldreich66}
\begin{equation}
 |\dot{E}_\mr{eq}|\sim\frac{E_\mr{tide}\Omega_\orb}{\Qeq},
\end{equation}
we have the approximate formula
\begin{equation}
 \Gdr \sim \frac{E_\mr{tide}}{\Qeq \mu a^2}\frac{k}{\omega},
\end{equation}
where the energy contained in the equilibrium tide is roughly $E_\mr{tide}\sim\lambda(M'/M)^2(R/a)^6\Es$, and the constant $\lambda$ (related to the apsidal motion constant) is $\sim3\times10^{-3}$ for an A star.

\citet{witte99} invoked the adiabatic approximation (\se{s:sec}) to show that resonance locking could occur in various fiducial eccentric binary systems;  \citet{burkart12} performed a similar analysis for KOI-54. However, as we have established in this work, this only means that the resonance locking fixed point exists, and not necessarily that resonance locking actually occurs. For the latter, we need to apply our criterion from \eq{e:capsucc}.

As in \se{s:wd}, we proceed to compute the values of our four parameters that affect the possibility of resonance locks in the current situation. We fix $r$, but assume $0\ll r<1$. We also assume that the star is nonrotating. We then find\footnote{Since we are considering an eccentric orbit, the tidal coupling coefficient $U$ (on which $\Gn$ depends) is additionally proportional to a Hansen coefficient, due to the Fourier expansion of the orbital motion (\se{s:modeamp}). We take this coefficient to be of order unity.}
\begin{align*}
 \frac{\gamma}{\omega} &= 10^{-11} \left( \frac{P_\orb}{40\ \mr{day}} \right)^{5/3}\\
 \frac{\Gdr}{\omega} &= 3\times10^{-21}\left( \frac{P_\orb}{40\ \mr{day}} \right)^{-4} \left( \frac{\Qeq}{10^8} \right)^{-1}\\
 \frac{\Gn}{\omega} &= 6\times10^{-18} \left(\frac{P_\orb}{40\ \mr{day}}\right)^{-4.6}.
\end{align*}

Substituting into \eq{e:capsucc}, noting that the first term on the right-hand side of \eq{e:capsucc} is much smaller than the second in this case (unlike in \se{s:wd}), we find that locks are present if
\begin{equation}
 1 > 10^{-4}\left( \frac{1-r}{0.1} \right)^{-1}\left( \frac{P_\orb}{40\ \mr{day}} \right)^{4.1}\left( \frac{\Qeq}{10^8} \right)^{-1}.
\end{equation}
It thus appears that resonance locks are indeed possible in eccentric binaries, subject to the validity of the assumptions enumerated in \se{s:assump} (e.g., solid-body rotation).

\section{Conclusion}\label{s:conc}
We have studied tidally induced resonance locking in close (but detached) binary systems. In a resonance lock, the detuning frequency $\domeg=\omega-\sigma$ between a stellar or planetary eigenmode frequency $\omega$ and a particular Fourier harmonic of the tidal driving frequency $\sigma=k\Omega_\orb-m\Omega_\spin$ is held constant (\se{s:basic}; \citealt{witte99}). This happens when a slowly varying physical process causing $\domeg$ to evolve in one direction is balanced by resonant interaction with the eigenmode in question causing $\domeg$ to evolve in the reverse direction.   The slowly varying process could be, e.g., orbital decay due to gravitational waves causing $\Omega_\orb$ to increase, magnetic braking causing $\Omega_\spin$ to decrease, stellar evolution altering $\omega$, or nonresonant components of the tidal response (the ``equilibrium tide'') affecting both $\Omega_\orb$ and $\Omega_\spin$ simultaneously.

Our primary goal has been to understand the dynamical properties and stability of resonance locks without relying on simplifying approximations for the mode amplitude evolution used in previous calculations. We defer detailed implications of these results to future papers. We have derived a novel set of  equations allowing for a dynamically evolving mode amplitude coupled to the evolution of both $\Omega_\spin$ and $\Omega_\orb$ (\se{s:form}). In particular, we do not assume that the mode amplitude is given by a Lorentzian profile resulting from the adiabatic approximation (\se{s:sec}) used in previous work, but instead solve the fully time-dependent mode amplitude equation.

In \se{s:rcap} we analyzed the stability of the dynamical fixed points associated with resonance locks. We analytically derived when such fixed points exist (\eqp{e:exist1}); there are either two fixed points or none for a given eigenmode. Although one of these equilibria is always unstable, the other can be stable when certain restrictions on binary and mode parameters are met (\eqp{e:estabul} in \se{s:stab}). One of the important conclusions of this analysis is that  resonance locks can exist and be stable even when the adiabatic approximation for the mode amplitude evolution is invalid (which happens, e.g., in the limit of moderately weak damping).

In \se{s:trajnum} we analyzed the  properties of resonance locks using direct numerical integration of our dynamical equations. In the simplest case in which a resonance lock fixed point exists and is stable, two possibilities arise: either a resonance passage is able to pump the mode's amplitude up sufficiently high to reach the fixed point and be captured into it, creating the resonance lock (\fig{f:trajplot1}), or the system instead sweeps through resonance without locking.

The more interesting situation is when both fixed points are unstable. In this case, we showed that resonance locking can nonetheless occur in some cases in a time-averaged sense. In these situations the mode amplitude and detuning frequency $\domeg$ execute limit cycles or even chaotic trajectories around the fixed points (see Figures \ref{f:trajplot2} \& \ref{f:trajplot4}). We presented evidence in \se{s:chaos} suggesting that resonance locking may in fact correspond to a strange attractor for certain parameter values; see \fig{f:chaosplot}.

In order to determine when resonance locking of some kind occurs (either stable, limit cycle, or chaotic), we performed numerical integrations over wide ranges of parameter values in \se{s:achievenum}. Using analytic approximations from \se{s:analytic}, we then developed approximate analytic formulae that explain our numerical results and define the binary and mode parameter regimes in which resonance locks of some kind occur. The key results are \eqs{e:capsuccret}{e:capsucc} and \fig{f:capregime}. Future studies of tidal evolution that do not include our full set of coupled mode-spin-orbit evolution equations can nonetheless utilize our results to assess whether resonance locks can occur.

One of the interesting consequences of resonance locks highlighted by \citet{witte99} and \citet{witte02} using numerical simulations in the adiabatic approximation is that locks can produce a significant speed up of orbital and spin evolution. We have explained this analytically in \se{s:amp}. In particular, we have demonstrated that resonance locks generically act to produce orbital and spin evolution on a timescale that is somewhat shorter than the slowly varying physical process whose influence drives the system into a lock. The magnitude of this acceleration depends on the effective moment of inertia ratio $r$ defined in \eq{e:r} and is large for $r \sim 1$. The latter condition can only be satisfied in eccentric binaries or binaries with high mass ratios. For the case of an eccentric orbit, we derived a rough condition for significant acceleration of the rate of orbital and spin evolution in \se{s:rval}.

To give a rough sense of the possible application of our results, we applied them to three sample astrophysical systems in \se{s:astroph}: inspiraling white dwarf and neutron star binaries in \se{s:wd}, and eccentric binaries with early-type stars in \se{e:ecc}. As has been argued previously using the adiabatic approximation for mode amplitudes, resonance locks are likely very common in white dwarf binaries and eccentric stellar binaries. They cannot, however, occur in neutron star binaries since orbital decay by gravitational wave emission is too rapid. A future application that may be of considerable interest is tidal circularization during high-eccentricity migration of hot Jupiters.

The theory of resonance locking that we have developed bears some similarity to resonance capture in planetary dynamics \citep{murray}. In the case of both mean-motion resonances \citep{goldreich65} and spin-orbit resonances \citep{goldreich68}, the generic equation governing the evolution of the relevant angle $\Psi$ towards resonance is
\begin{equation}\label{e:planres}
   \ddot{\Psi} =- F\sin\Psi + G.
\end{equation}
For mean-motion resonances, $\Psi$ defines the angle between the mean anomalies of two orbiting bodies, while for spin-orbit resonances, $\Psi$ is related to the difference between a body's mean and true anomalies (relevant only for eccentric orbits). In both cases, $G$ provides a frequency drift term, resulting from orbital decay for mean-motion resonances and from a net tidal torque for spin-orbit resonances.

\Eq{e:planres} can be compared to our equation describing the evolution of the detuning frequency $\domeg=\omega-\sigma$ in \eq{e:eomsig2}. Both specify the second time derivative of a resonance angle, and both contain a frequency drift term describing how resonance is approached. The essential difference is that in place of the pendulum restoring force present in \eq{e:planres}, tidal resonance locking instead contains two terms providing the complicated interaction with a stellar or planetary normal mode. Thus, although there are qualitative similarities between resonance capture in planetary dynamics and resonance locking, no formal mathematical analogy exists.

It is important to reiterate that resonance locks can only occur under a specific set of conditions (\se{s:assump}). They are not relevant to all close binaries. In particular, the dynamical tide must be composed of global radial standing waves, with damping times much longer than radial wave travel times. For this reason, an efficient angular momentum transport process must maintain approximate solid body rotation; if not, critical layers may develop where mode angular momentum is deposited into the background rotation profile, which would lead to efficient local wave damping. In addition, we have restricted our analysis to linear perturbation theory. In practice, this represents a restriction on the maximum mode amplitudes that are allowed, since nonlinear instabilities can act on large-amplitude waves. For example, in the case of binaries containing solar-type stars with radiative cores, wave breaking in the core likely prohibits the establishment of global standing waves \citep{goodman98}, thus also precluding resonance locks from developing.

\section*{Acknowledgements}
The authors are pleased to thank Edgar Knobloch, Keaton Burns, Eugene Chiang, and Scott Tremaine for useful discussions and guidance. J.B. is an NSF Graduate Research Fellow. E.Q. was supported by a Simons Investigator award from the Simons Foundation, the David and Lucile Packard Foundation, and the Thomas Alison Schneider Chair in Physics at UC Berkeley. P.A. is an Alfred P. Sloan Fellow, and received support from the Fund for Excellence in Science and Technology from the University of Virginia.

\appendix

\section{Deriving fixed point stability conditions}\label{a:fixstab}
Here we will determine the stability region for the resonance locking fixed point described in \se{s:rcap}. The characteristic polynomial $P$ in question is given in \eq{e:char}.

First, $P_1>0$ reduces to
\begin{equation}
 \domegf < \frac{(1-r)^2\gamma^2\Gn}{r\Gdr^2}.
\end{equation}
We immediately see that this is satisfied if $\Gdr<0$ (assuming $\Gn>0$), since by \eq{e:rightfp} we then have $\domegf<0$. If instead $\domegf>0$, then we can take $|\domegf|\gg\gamma$ to derive
\begin{equation}\label{e:ineqrep}
   \Gdr<\gamma\\\left( \frac{1-r}{r^{2/3}} \right)\left( \frac{\Gn}{\omega} \right)^{1/3}.
\end{equation}

Next, the Hurwitz matrix for $P$ is
\begin{equation}
   H=\begin{pmatrix}
      2\gamma & P_0 \\
      1 & P_1 \\
        & 2\gamma & P_0
   \end{pmatrix}.
\end{equation}
The remaining condition for stability of the fixed point is that the three leading principal minors of $H$ must be positive; this formally yields three additional inequalities. However, one is $\gamma>0$ which is always satisfied, and the other two are actually identical:
\begin{equation}
 2\gamma P_1 > P_0,
\end{equation}
which expands to
\begin{equation}
 \gamma^2\Gn(1-r)^2 > \Gdr^2\domegf.
\end{equation}
We now see that $\Gdr<0$ implies asymptotic stability. If $\Gdr>0$, then the inequality reduces to (again assuming $|\domegf|\gg\gamma$)
\begin{equation}\label{e:stabapp}
   \Gdr<\gamma\left( 1-r \right)\left( \frac{\Gn}{\omega} \right)^{1/3}.
\end{equation}
Since \eq{e:stabapp} is more restrictive than \eq{e:ineqrep} (since $r<1$ for $\Gdr>0$; see \eqp{e:exist2}), \eq{e:stabapp} is the condition for asymptotic stability when $\Gdr>0$.

\section{Hansen coefficient scaling}\label{a:hansen}
Here we will determine how the Hansen coefficients $X_{lm}^k$ scale with harmonic index $k$. These coefficients are defined to satisfy
\begin{equation}
 \left( \frac{a}{D(t)} \right)^{l+1}e^{-imf(t)}=\sum_{k=-\infty}^\infty X_{lm}^k e^{-ik\Omega_\orb t},
\end{equation}
where $D$ is the binary separation and $f$ is the true anomaly. For $|k|\gg |m|\Omega_\mr{peri}/\Omega_\orb$, where $\Omega_\mr{peri}$ is the effective orbital frequency at periapse, we have that $X_{lm}^k\propsim X_{00}^k$. We can express $X_{lm}^k$ in general as an integral over the eccentric anomaly $E$ \citep{burkart12}; for $l=m=0$, this is
\begin{equation}\begin{split}
 X_{00}^k &= \frac{1}{\pi}\int_0^\pi \cos\left[ k(E-e\sin E) \right] dE\\
 &= J_k(ek)
\end{split}\end{equation}
where $J$ is a Bessel function. We can expand $J_k(ek)$ as \citep{abramowitz}
\begin{equation}
  J_k(ek) \propsim \frac{1}{\sqrt{k}} \exp\left[ -k g(e)\right],
\end{equation}
where
\begin{equation}\begin{split}\label{e:g}
 g(e) &= \frac{1}{2}\ln\left( \frac{1+\eta}{1-\eta} \right)-\eta\\
 &= \frac{\eta^3}{3} + \frac{\eta^5}{5} + \frac{\eta^7}{7} + \cdots
\end{split}
\end{equation}
and $\eta = \sqrt{1-e^2}$.

\section{Canonical angular momentum}\label{a:J}
Here we will derive the canonical angular momentum associated with a stellar perturbation. The Lagrangian density for a stellar perturbation is \citep{friedman78a}
\begin{equation}
 \mathcal{L} = \frac{1}{2}\rho\left( |\dot{\vec{\xi}}|^2 + \dot{\vec{\xi}}\cdot B\vec{\xi} -  \vec{\xi}\cdot C\vec{\xi}\right),
\end{equation}
where $\vec{\xi}$ is the Lagrangian displacement vector, the Coriolis force operator $B$ was defined in \se{s:modeamp}, and the Hermitian operator $C$ is defined in e.g.\ \citet{schenk02}. The $z$ component of the canonical angular momentum is then
\begin{equation}\begin{split}
 J &=  -\left\langle\partial_\phi\vec{\xi},\,\frac{\partial\mathcal{L}}{\partial\dot{\vec{\xi}}}\right\rangle\\
 &= -\left\langle\partial_\phi\vec{\xi},\, \dot{\vec{\xi}}+B\vec{\xi}/2\right\rangle.
\end{split}\end{equation}

As in \se{s:modeamp}, we perform a phase space expansion of the Lagrangian displacement vector and its time derivative \citep{schenk02}, so that
\begin{equation}
 \begin{pmatrix}
  \vec{\xi}\\\dot{\vec{\xi}}
 \end{pmatrix}
 = \sum_A q_A
 \begin{pmatrix}
  \vec{\xi_A}\\-i\omega_A\vec{\xi_A}
 \end{pmatrix},
\end{equation}
where $A$ runs over all rotating-frame stellar eigenmodes and both frequency signs. The canonical angular momentum then becomes
\begin{equation}\label{e:modeJ1}
 J = \frac{1}{2} \sum_{AB} q_A^* q_B^{}\, m_A^{} \left[ 2\omega_B\left\langle\vec{\xi}_A,\vec{\xi}_B\right\rangle + \left\langle\vec{\xi}_A,iB\vec{\xi}_B\right\rangle\right].
\end{equation}
Since $\left\langle\partial_\phi\vec{\xi},\, \dot{\vec{\xi}}\right\rangle$ is real valued, we can re-express \eq{e:modeJ1} as
\begin{equation}
 J = \frac{1}{2} \sum_{AB} q_A^* q_B^{} \left[ (m_A\omega_B+m_B\omega_A)\left\langle\vec{\xi}_A,\vec{\xi}_B\right\rangle + m_A\left\langle\vec{\xi}_A,iB\vec{\xi}_B\right\rangle\right].
\end{equation}
We have that $\left\langle\vec{\xi}_A,\vec{\xi}_B\right\rangle \propto \delta_{m_A,m_B}$, so $J$ reduces to
\begin{equation}
 J = \frac{1}{2} \sum_{A} \frac{\epsilon_A m_A}{\omega_A}|q_A|^2,
\end{equation}
where we have used the orthogonality relation \citep{friedman78b}
\begin{equation}
 (\omega_A+\omega_B)\left\langle\vec{\xi}_A,\vec{\xi}_B\right\rangle + \left\langle\vec{\xi}_A,iB\vec{\xi}_B\right\rangle = \delta_{AB} \frac{\epsilon_A}{\omega_A}.
\end{equation}

\bibliography{rcap}

\end{document}